\documentclass[twocolumn,usenames,dvipsnames]{aastex63}

\usepackage{graphicx}	
\usepackage{caption,subcaption}
\usepackage{amsmath}	
\usepackage{amssymb}	
\usepackage{relsize}
\usepackage{mathtools}
\usepackage{bigints}

\usepackage{floatrow}
\usepackage{ulem}
\usepackage{soul}
\usepackage{comment}
\usepackage{bm}

\usepackage{macros_vdb}
\graphicspath{{./}{figures/}}

\received{XXX}
\revised{YYY}
\accepted{ZZZ}

\begin{document}

\shorttitle{Dynamical Friction and Core Stalling -- I.}
\shortauthors{Banik and van den Bosch}

\title{Dynamical Friction, Buoyancy and Core-Stalling -- I. A Non-perturbative Orbit-based Analysis}

\correspondingauthor{Uddipan Banik}
\email{uddipan.banik@yale.edu}

\author[0000-0002-9059-381X]{Uddipan Banik}
\affiliation{Department of Astronomy, Yale University, PO. Box 208101, New Haven, CT 06520, USA}

\author[0000-0003-3236-2068]{Frank~C.~van den Bosch}
\affiliation{Department of Astronomy, Yale University, PO. Box 208101, New Haven, CT 06520, USA}

\label{firstpage}


\begin{abstract}

  We examine the origin of dynamical friction using a non-perturbative, orbit-based approach. Unlike the standard perturbative approach, in which dynamical friction arises from the LBK torque due to pure resonances, this alternative, complementary view nicely illustrates how a massive perturber significantly changes the energies and angular momenta of field particles on near-resonant orbits, with friction arising from an imbalance between particles that gain energy and those that lose energy.  We treat dynamical friction in a spherical host system as a restricted three-body problem. This treatment is applicable in the `slow' regime, in which the perturber sinks slowly and the standard perturbative framework fails due to the onset of non-linearities. Hence it is especially suited to investigate the origin of core-stalling: the cessation of dynamical friction in central constant-density cores. We identify three different families of near-co-rotation-resonant orbits that dominate the contribution to dynamical friction. Their relative contribution is governed by the Lagrange points (fixed points in the co-rotating frame). In particular, one of the three families, which we call \pacman orbits because of their appearance in the co-rotating frame, is unique to cored density distributions. When the perturber reaches a central core, a bifurcation of the Lagrange points drastically changes the orbital make-up, with \pacman orbits becoming dominant. In addition, due to relatively small gradients in the distribution function inside a core, the net torque from these \pacman orbits becomes positive (enhancing), thereby effectuating a dynamical buoyancy. We argue that core stalling occurs where this buoyancy is balanced by friction.
\end{abstract}


\section{Introduction}
\label{sec:intro}

Dynamical friction is an important relaxation mechanism in gravitational $N$-body systems like galaxies and clusters. Massive objects such as black holes, globular clusters and dark matter subhaloes lose energy and angular momentum to the field particles and sink to the centers of their host systems, driving the system towards equipartition. \cite{Chandrasekhar.43} was the first to derive an expression for the dynamical friction force on a massive object (hereafter the `perturber') travelling through a homogeneous medium on a straight orbit, by summing the velocity changes from independent two body encounters with the field particles. Despite its obvious over-simplifications, applying the formula for Chandrasekhar's friction force using the {\it local} density and velocity distribution of the particles in an {\it inhomogeneous} body, such as a halo or galaxy, yields results that are in fair agreement with numerical simulations \citep[][]{Lin.Tremaine.83, Cora.etal.97, vdBosch.etal.99, Hashimoto.etal.03, Boylan-Kolchin.etal.08, Jiang.etal.08}. However, this `local approximation' fails to account for the cessation of dynamical friction in the central regions of halos or galaxies with a constant-density core. This so-called core-stalling has been observed in $N$-body simulations \citep[e.g.,][]{Read.etal.06c, Inoue.11,  Petts.etal.15, Petts.etal.16, DuttaChowdhury.etal.19} but is still not properly understood. In addition, the simulations also show that prior to stalling the object often experiences a short phase of enhanced `super-Chandrasekhar friction', followed by a `kick-back' effect in which it is pushed out before it settles at the `core-stalling radius' \citep[][]{Goerdt.etal.10, Read.etal.06c, Zelnikov.Kuskov.16}. In fact,  \citet{Cole.etal.12} have shown that massive objects initially placed near the center of a cored galaxy experience a `dynamical buoyancy' that pushes them out towards this stalling radius. This complicated phenomenology cannot be explained using Chandrasekhar's treatment of dynamical friction, which instead predicts that the orbits of massive objects continue to decay inside a central core region, albeit at a reduced rate \citep[e.g.,][]{Hernandez.Gilmore.98, Banik.vdBosch.2021}. 

Dynamical buoyancy can have important astrophysical implications in cored galaxies, where it can either push out massive objects such as nuclear star clusters and supermassive black holes from the central regions, or stall their in-fall (core-stalling) by counteracting the effect of dynamical friction. The latter has been invoked by \cite{Goerdt.etal.10} and \cite{Cole.etal.12} to explain the survival of the globular clusters in the Fornax dwarf galaxy, hinting at the possibility of a central dark matter core.

Given that Chandrasekhar's expression for the dynamical friction force is based on the highly idealized assumption of straight orbits in a uniform, isotropic background, it should not come as a surprise that there are circumstances under which it fails. \citet[hereafter TW84]{Tremaine.Weinberg.84} generalized the description of dynamical friction to a more realistic system of an inhomogeneous spherical galaxy with a small, time-dependent perturbation (bar or satellite). Using Hamiltonian perturbation theory to perturb the actions of the field particles (or `stars') up to second order in the perturbation parameter, they infer that dynamical friction arises from a net retarding torque on the perturber from stars along purely resonant orbits (whose orbital frequencies are commensurable with the circular frequency of the perturber). This torque, known as the LBK torque, was first derived by \cite{LyndenBell.Kalnajs.72} in the context of angular momentum transport driven by spiral arms in disk galaxies. \citet[hereafter KS18]{Kaur.Sridhar.18} showed that for a cored \cite{Henon.59} Isochrone galaxy the LBK torque vanishes at a certain radius in the core due to the suppression in the number of contributing resonances and reduction of the strength of the torque from the surviving resonances, causing the perturber to stall. However their treatment does not explain the origin of super-Chandrasekhar dynamical friction or dynamical buoyancy.

In \citet[hereafter BB21]{Banik.vdBosch.2021}, we showed that an exclusive contribution from resonances between the perturber and the field particles to the LBK torque, as obtained by TW84 and KS18, is ultimately a consequence of two key assumptions, the adiabatic (slow growth of the perturber) and secular (slow in-fall under dynamical friction) approximations, which effectively boil down to ignoring the effect of friction-driven in-fall in the computation of the torque. In BB21 we relaxed these two assumptions and properly accounted for the time dependence of the location and circular frequency of the perturber (due to its radial in-fall motion) to compute the response density and the corresponding self-consistent torque, $\calT_{\rm SC}$. This differs from the standard LBK torque in two key aspects: (i) it has a significant contribution from near-resonant orbits, and (ii) it not only depends on the instantaneous orbital radius of the perturber, $R(t)$, but on its entire in-fall history by involving a temporal correlation of the perturber potential. We showed that super-Chandrasekhar dynamical friction, dynamical buoyancy and core-stalling can all be explained as consequences of this ``memory effect".

Although this self-consistent formalism is more general than the standard LBK formalism and offers predictions related to core stalling that qualitatively match those from numerical simulations, it suffers from a few caveats. First of all, in order to avoid having to solve the complicated integro-differential equation for the self-consistent evolution of $R(t)$, BB21 assume the in-fall rate, $\rmd R/\rmd t$, to be slowly varying over time. This allows $\calT_{\rm SC}$ to be written as the sum of an instantaneous torque, $\calT_{\rm inst}$, that depends on time $t$ and the orbital radius $R(t)$, and a memory torque, $\calT_{\rm mem}$, that is proportional to $\rmd R/\rmd t$. The latter becomes dominant in the core region and acts as a source of destabilizing feedback, giving rise to an accelerated super-Chandrasekhar in-fall outside a critical radius, $\Rcrit$. Inside $\Rcrit$, the memory torque flips sign and becomes enhancing, i.e., exerts dynamical buoyancy. The perturber is thus found to stall at $\Rcrit$ due to a balance between friction outside and buoyancy within, i.e., $\Rcrit$ acts as an attractor. However, the critical behaviour near this radius ($\rmd R/\rmd t \to \pm \infty$ as $R\to \Rcrit$ instead of approaching zero as is typical for a stable attractor) is an artefact of the assumption of a near-constant $\rmd R/\rmd t$, which becomes questionable close to $\Rcrit$ as the perturber undergoes an accelerated in-fall before stalling at this radius. This critical behaviour can be smoothed out by solving the integro-differential equation for $R(t)$ in its full generality, which is however a non-trivial exercise. 

The second caveat of the self-consistent formalism (and of previous studies like TW84 and KS18) is related to the concept of resonances in linear perturbation theory. In this perturbative picture, dynamical friction is driven by resonances between the unperturbed frequencies of the stars and the perturber. But these resonances themselves drastically change (`perturb') the actions and frequencies of the resonant stars, questioning the very assumption of a weak perturbation. TW84 address this philosophical issue by introducing the concept of `sweeping through the resonances', i.e., linear perturbation theory only holds in the `fast' regime, where the circular frequency of the perturber changes rapidly under dynamical friction such that the stars fall out of resonance before their actions can change significantly and give rise to non-linear perturbations in the distribution function. However, in a cored galaxy, as the perturber slows down upon approaching the stalling radius, stars no longer sweep fast enough through the resonances. Therefore, perturbation theory, especially a linear order one, becomes questionable in this `slow' regime. 

The final caveat relates to the fact that linear perturbation theory assumes a weak perturbing potential, i.e., the mass of the perturber, $M_\rmP$, is much smaller than the galaxy mass enclosed within $R$, $M_\rmG(R)$. Numerical simulations, though, have shown that near the stalling radius $M_\rmG(R)$ is actually comparable to $M_\rmP$ \citep[][]{Petts.etal.15, Petts.etal.16, DuttaChowdhury.etal.19}, indicating that the torque is likely to have an appreciable contribution from non-linear perturbations in the distribution function.

Simply put, then, linear perturbation theory is inadequate to describe the dynamics related to core stalling. In order to overcome this conceptual problem, in this paper we develop a {\it non-perturbative} formalism to investigate how dynamical friction operates in the `slow' regime, i.e., near the core stalling radius. We adopt a circular restricted three body framework and integrate the orbits of massless field particles in the combined potential of a host galaxy and a massive perturber (to arbitrary order) moving along a circular orbit. We find that the dominant contribution to the torque comes from a family of \textit{near-co-rotation-resonant} orbits that slowly drift (librate) around the Lagrange points in the co-rotating frame. The nature of these orbits is found to change drastically as one approaches the core region of a galaxy. This causes a transition from a state in which the majority of orbits cause a retarding torque on the perturber (`dynamical friction'), to one in which the torque becomes predominantly enhancing (`dynamical buoyancy'). As discussed in Paper II in this series (Banik \& van den Bosch, in preparation), this transition is associated with a bifurcation in the Lagrange points that occurs whenever the perturber reaches a  characteristic radius, $R_{\rm bif}$, which we associate with the core stalling radius.

This paper is organized as follows. In Section~\ref{sec:concept} we first conceptualize, without resorting to mathematics, how dynamical friction on a massive perturber arises from a net torque exerted by particles on near-co-rotation-resonant orbits. We then introduce, in Section~\ref{sec:threebody}, the restricted three-body framework used throughout this paper. In Section~\ref{sec:orbits} we introduce the various orbital families that arise in the presence of a massive perturber, and briefly discuss how they contribute to dynamical friction. In Section~\ref{sec:resonances} we describe a non-perturbative method to compute the integrated energy and angular momentum transfer from individual orbits, and show that certain orbital families in a cored galaxy can give rise to a positive, enhancing torque (dynamical buoyancy) in the core region, the origin of which we examine in Section~\ref{sec:core}. We summarize our findings in Section~\ref{sec:concl}.

\begin{figure*}
\includegraphics[width=1\textwidth]{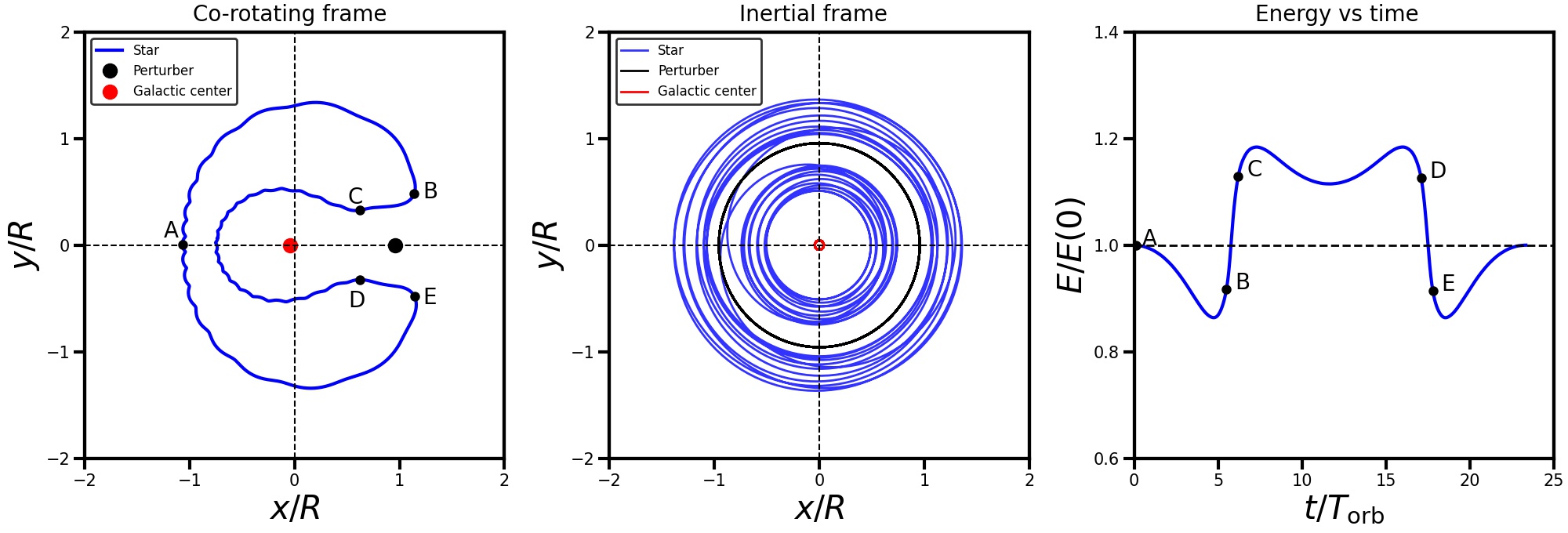}
  \caption{\small Example of a NCRR \horseshoe orbit. The left-hand panel shows the orbit in the co-rotating frame, in which the perturber (indicated by a thick, solid black dot) is at rest at $(x,y) = (R,0)$. The red dot marks the center of the galaxy, while the letters A,B,..,E mark specific points along the orbit. The middle panel shows the same orbit, but now in the inertial frame. Note how the orbit librates back and forth between regions inside and outside of the perturber. The right-hand panel depicts how a field particle moving along this \horseshoe orbit changes its orbital energy with time. Because of the near-co-rotation resonance nature of this orbit, it takes many orbital periods of the perturber, $T_{\rm orb}$, to complete one \horseshoe (in this case, the libration time $T_{\rm lib} \sim 24\, T_{\rm orb}$). The largest energy changes occur when the field particle moves from outside of the perturber (outer section) to inside (inner section), and vice-versa, which corresponds to the transitions from B to C and from D to E, respectively.}
  \label{fig:horseshoe}
\end{figure*}

\section{Conceptualizing Dynamical Friction}
\label{sec:concept}

The non-perturbative framework adopted here gives an alternative, complementary view of dynamical friction, which is subtly different from the standard resonance picture presented in TW84 and KS18. In this section we conceptualize this alternative view using the example of a single orbit. Without going into any mathematical detail, which is relegated to Sections~\ref{sec:threebody}-\ref{sec:core}, the goal is to illustrate, in a pictorial view, how dynamical friction arises. This serves to underscore the complicated, higher-order nature of dynamical friction, and to hopefully clarify the more technical treatment that follows.

As we do throughout this paper, we consider a massive body, the perturber, orbiting a large system (hereafter the galaxy) consisting of a large number, $N$, of `field' particles or stars. Throughout, we  simplify the picture by assuming that both perturber and galaxy are spherically symmetric, and that the perturber is on a planar, circular orbit within the galaxy at a galacto-centric radius $R$. We assume that the mass of the field particles, $m$, is negligible compared to that of either the perturber, $M_\rmP$, or the galaxy, $M_\rmG$. In addition, we ignore the radial motion of the perturber due to dynamical friction/buoyancy, since we are interested in the dynamics near the stalling radius. Hence, we can treat our dynamical system as a circular restricted three body problem, which dramatically simplifies the dynamics since the gravitational potential is now static in the frame co-rotating with the perturber. Here, and throughout this section, we assume an isotropic \cite{Plummer.11} galaxy and a point mass perturber with a mass that is 0.4 percent of the galaxy mass on a circular orbit at half the scale radius of the galaxy.
 
As we discuss in Section~\ref{sec:orbits}, one can distinguish a number of different orbital families in the co-rotating frame. Here we focus on one example; the \horseshoe orbit, which, as we will show, is one of the key actors in our dynamical friction narrative. Fig.~\ref{fig:horseshoe} shows an example of a \horseshoe orbit, both in the co-rotating frame (left-hand panel), in which it takes on a shape to which it owes its name, and in the inertial frame (middle panel). A field particle on this orbit is in near-co-rotation resonance (hereafter NCRR) with the perturber in that the azimuthal frequency, $\Omega_\phi$, with which it circulates the center of the unperturbed galaxy is very similar to that of the perturber's circular orbit, $\Omega_\rmP$. Since we assume that the perturber orbits in the anti-clockwise direction, all orbits in the co-rotating frame will have a net clockwise drift motion around their center of circulation. The NCRR orbits librate about the Lagrange points and are therefore often called `trapped' orbits \citep[e.g.,][]{Barbanis.76, Sellwood.Binney.02, Daniel.Wyse.15, Contopoulos.73, Contopoulos.79, Goldreich.Tremaine.82}. However, since many of these orbits are not strictly trapped, in that they often undergo separatrix crossings (see Section~\ref{sec:orbfam} below), we consider the nomenclature NCRR more explicit.

Let us assume that the field particle starts out at position A (indicated in the left-hand panel of Fig.~\ref{fig:horseshoe}) on the \horseshoe orbit. Since it is farther away from the center-of-mass than the perturber, it circulates slower. Slowly, with an angular speed of roughly $\Omega_\rmP - \Omega_\phi$, the perturber catches up with the field particle, coming closer and closer. In the co-rotating frame, this corresponds to the field particle travelling upwards, clockwise, along its orbit. As it slowly librates from A ($t=0$) to B, its energy and angular momentum increase (note the gradual decrease in $E/E(0)$ from A to B in the right-hand panel of Fig.~\ref{fig:horseshoe}). When it reaches point B, the perturber exerts an inward accelerating force, pulling the particle onto the inner, more bound arc of the orbit. As the particle moves from B to C, it crosses co-rotation resonance; its orbital energy decreases steeply and its azimuthal frequency, $\Omega_\phi$, now becomes larger than $\Omega_\rmP$. Note that, since the Hamiltonian of our perturbed system is time-variable, energy is not a conserved quantity (and neither is angular momentum nor $\Omega_\phi$). However, the total energy of the system is conserved, and the energy that the field particle loses as it transits from B to C is transferred to the perturber, which will move (very slightly) outward; this is the opposite of dynamical friction, to which we refer as dynamical buoyancy. 

Once the field particle arrives at C, the particle now circulates {\it faster} than the perturber, and it starts to drift farther and farther ahead of the perturber (in the co-rotating frame). It circulates around the center of the galaxy (as we will see below, it has to go all the way around the center because of the potential barrier associated with an unstable Lagrange point, or saddle, in between the perturber and the center), and ultimately makes its way to point D, where the perturber exerts an outward pulling force, which puts the particle back on the outer arc of its orbit. This time, the perturber gives energy to the field particle, thus experiencing dynamical friction. Once at point E, the particle starts to lag behind the perturber again, until it drifts back to (close to) its original position A. 

In the restricted three-body problem considered here, the Jacobi energy, unlike the orbital energy, is a conserved quantity (see Section~\ref{sec:threebody}). This ensures that the energy gain experienced by the perturber at $B \rightarrow C$ balances the energy loss experienced at $D \rightarrow E$. In other words, the {\it net} effect on the perturber of a field particle along this NCRR orbit is zero. 

\begin{figure*}[t!]
  \centering
  \includegraphics[width=1\textwidth,height=0.27\textwidth]{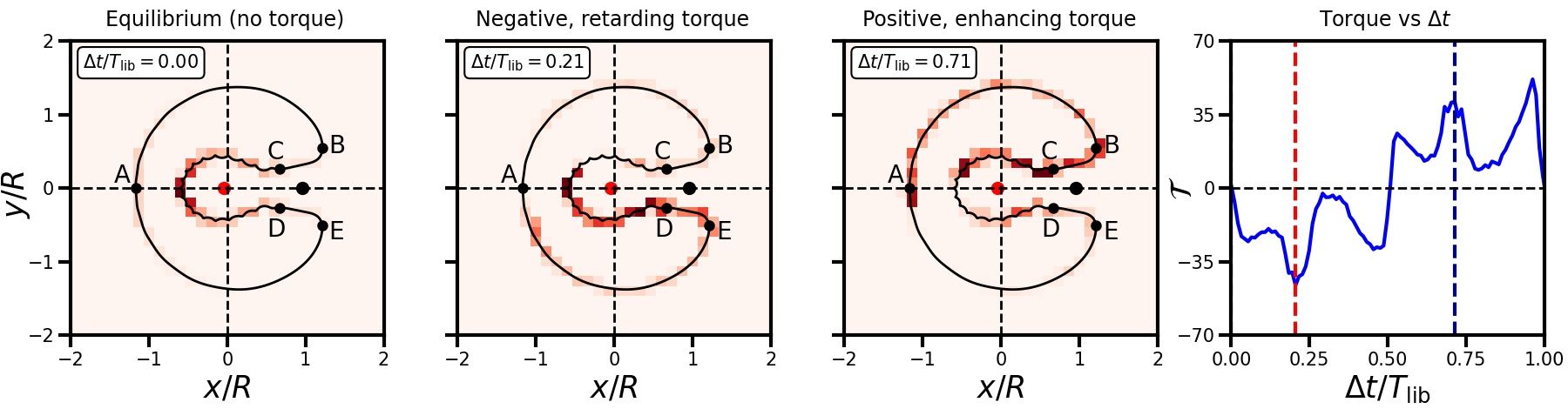}
  \caption{Illustration of the origin of torque on the perturber from a NCRR orbit. The heat maps show the distribution of field particles in the co-rotating frame along a \horseshoe orbit as in Fig.~\ref{fig:horseshoe}, with darker colors indicating a larger number density. The rightmost panel shows the evolution of the torque (as a function of time in units of $T_{\rm lib}$, the libration time or the time taken for $2\pi$ circulation in the co-rotating frame) as the field particles move along the orbit. At $\Delta t=0$ (first panel), the unperturbed density distribution of field particles is spherically symmetric, and there is no net torque on the perturber. However, some time later (second panel, corresponding to $\Delta t$ marked by the red dashed line in the right-most panel), the particles have shifted along the orbit, resulting in an enhanced density of field particles lagging behind the perturber, giving rise to a retarding torque. If the perturber would remain on its original orbit, then some time later (many orbital periods since the drift/libration time along the \horseshoe is long) the particles would have drifted to the location depicted in the third panel (at $\Delta t$ marked by the blue dashed line in the rightmost panel), exerting an enhancing torque exactly opposite to that depicted in the second panel. When integrating over the entire libration period, the net torque is therefore zero. Dynamical friction arises only because the initial torque is retarding, after which the perturber moves in, and the near-resonant frequencies change (i.e., one never makes it to the point shown in the third panel).}
  \label{fig:wake}
\end{figure*}

So how, then, does dynamical friction arise? The two key ingredients that give rise to net dynamical friction are the long libration (or `drift') time of these NCRR orbits, and the non-uniform density distribution of field particles as a function of orbital phase. The libration time, $T_{\rm lib}$, is the time in which the field particle completes a full \horseshoe (i.e., from $A \rightarrow B \rightarrow C \rightarrow D \rightarrow E \rightarrow A$). Because the orbit is in near-co-rotation resonance, this is much longer than the orbital periods of the perturber or the field particle. The non-uniform distribution of particles along the orbit can be understood as follows: in the limit of large $N$, there are many field particles that are on the same (or at least on a very similar) orbit. All these particles have different orbital phases, though. Consider the unperturbed galaxy, which is assumed to be in equilibrium and characterized by a distribution function $f_0(\bx,\bv)$. This unperturbed distribution function determines how many field particles are mapped onto each phase of each orbit once the perturber is introduced (here, for the sake of simplicity, we assume that the perturber is introduced instantaneously). Typically, since the density increases towards the center, the number density of particles on the inner arc of the \horseshoe ($C-D$) is larger than along the outer arc ($E-A-B$). This is depicted in the left-most panel of Fig.~\ref{fig:wake}, where darker colors indicate a larger number density of field particles. These have been computed using the (isotropic) distribution function of our (unperturbed) Plummer sphere, under the assumption that this captures the distribution of field particles along this orbit at time $t=0$, when the perturber is introduced. Some time $\Delta t < T_{\rm lib}$ later, all the particles have drifted along the \horseshoen, and the phase-dependent number density distribution now looks similar to that in the second panel: because of the initial non-uniformity in orbital phases, there are now more particles along the $D \rightarrow E$ part of the orbit than along the $B \rightarrow C$ part; there are more energy gainers than energy losers, causing a net energy loss of the perturber. Or, in terms of angular momentum, the overdensity of field particles trailing the perturber, exerts a torque that reduces the perturber's orbital angular momentum (note the negative, retarding torque at this time, marked by the red dashed line in the rightmost panel that shows the evolution of the torque exerted by the particles). Hence, during this phase of the evolution, the perturber experiences (net) dynamical friction from the field particles associated with this \horseshoe orbit.

If the perturber would remain at its current orbital radius (i.e., if we temporarily ignore the consequences of dynamical friction), then the phase of the overdensity of particles along the \horseshoe orbit would continue to drift around, ultimately making its way to points $B$ and $C$ (depicted in the third panel of Fig.~\ref{fig:wake}), where it would exert a positive, enhancing torque/ buoyancy on the perturber (marked by the blue dashed line in the rightmost panel) which nullifies the initial dynamical friction on the perturber\footnote{The alternating phases of retarding and enhancing torques from the NCRR orbits are responsible for oscillations in the pattern speed of a galactic bar in the slow regime of dynamical friction, as noted by \cite{Chiba.Schonrich.21}.}. However, because of the long drift time, the time between this net friction and equal, but opposite, net buoyancy is very long ($\sim 10\, T_{\rm orb}$ for the specific \horseshoe orbit shown in Fig.~\ref{fig:wake}). 
During this time, the initial net friction from many NCRR orbits will have caused the perturber to move inward, to a more bound orbit. This changes its orbital frequency, $\Omega_\rmP$, such that, by the time the overdensity {\it would} have reached point $B$, the system has changed sufficiently that new field particles have now entered near-co-rotation resonance with the perturber and those associated with our original \horseshoe orbit have fallen out of resonance. Dynamical friction is therefore a secular process; the field particles drain energy from the perturber, causing it to in-fall, which in turn changes the orbital frequencies, facilitating further energy transfer. This process of `sweeping through the resonances' by the perturber is crucial for dynamical friction to operate, as emphasized in great detail in TW84. 

\subsection{The Role of Resonances}

In the perturbative framework of TW84 and KS18, dynamical friction arises from the LBK torque which only has a non-zero contribution from pure resonances, i.e., orbits that obey a commensurability condition between the (circular) frequency of the perturber, $\Omega_\rmP$, and the frequencies of the field particles in the {\it unperturbed} potential. Even the more general, self-consistent torque introduced by BB21, is formulated in terms of these frequencies. 

In the non-perturbative framework adopted in this paper, in which we consider fully perturbed orbits\footnote{To clarify the paradoxical use of `perturbed orbits in a `non-perturbative framework'; perturbative is used to mean `as pertaining to perturbation theory', whereas perturbed means `impacted by the in-falling, perturbing mass'.} in the galaxy$+$perturber potential to arbitrary order, the frequencies of the individual field particles vary with time due to energy and angular momentum exchanges with the perturber; the original actions of the unperturbed galaxy are no longer conserved, and neither are the frequencies associated with the corresponding angles \citep[][]{Tremaine.Weinberg.84,Fouvry.Bar-Or.18}. Hence, a field particle will not satisfy a commensurability condition throughout its orbital evolution but rather will find itself `trapped', librating around resonance(s) with the perturber. In fact, this is what happens when the field particle along the \horseshoe orbit in Fig.~\ref{fig:horseshoe} moves from B to C and from D to E; it's azimuthal frequency, $\Omega_\phi$, is swept back and forth through a near-co-rotation resonance with the circular frequency of the perturber,  $\Omega_\rmP$. This same principle also underlies the physics of radial migration in disks due to interactions with transient spirals \citep[e.g.,][]{Carlberg.Sellwood.85, Sellwood.Binney.02, Daniel.Wyse.15}. Dynamical friction arises from an imbalance between the number of field particles that `sweep up' versus `sweep down' in frequency space, and this imbalance itself arises from gradients in the distribution function.

\section{The Restricted Three Body Problem}
\label{sec:threebody}

We treat dynamical friction as a restricted three-body problem, in which the mass of the field particles is negligible compared to that of the galaxy and the perturber. Throughout, we assume that both galaxy and perturber are spherically symmetric, and that the perturber is moving along a circular orbit of galacto-centric radius $R$ within the galaxy. In this setting the gravitational potential is static (in the absence of dynamical friction) in the co-rotating frame, which greatly simplifies the analysis that follows. As the perturber only feels the gravitational field of the galaxy mass enclosed within a sphere of radius $R$ centered on the galactic center, denoted by $M_\rmG(R)$, we follow \cite{Inoue.11} and KS18 in assuming that $M_\rmP$ and $M_\rmG(R)$ rotate about their common center of mass (hereafter COM).

\begin{figure}
\includegraphics[width=0.9\textwidth,height=0.89\textwidth]{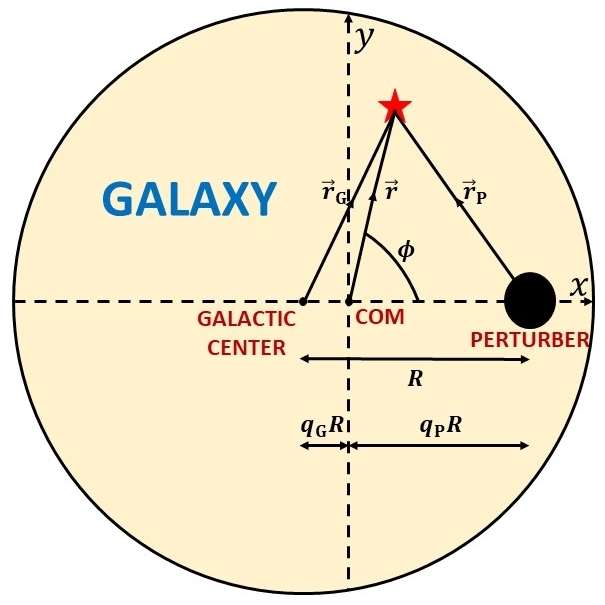}
\caption{\small Schematic of a massive perturber on a circular orbit in a spherically symmetric galaxy. The co-rotating $(x,y)$-frame is centered on the COM with the $x$ axis pointing in the direction of the perturber.}
\label{fig:schematic}
\end{figure}

\subsection{Models for the Galaxy and the Perturber}
\label{sec:models}

The geometry of our dynamical model is illustrated in Fig.~\ref{fig:schematic}. It depicts the galaxy (large, shaded circle), the perturber (solid black dot), and the COM in the co-rotating $(x,y)$-frame that we will adopt throughout. For convenience, we define the following mass ratios: $q \equiv M_\rmP/M_\rmG$ is the mass ratio of the in-falling perturber and the host galaxy, while $q_{\rm enc}(R) \equiv M_\rmP/M_\rmG(R)$ is the mass ratio of the perturber and the galaxy enclosed within $R$. The distances between the COM and the galactic center and between the COM and the perturber are given by $q_\rmG R$ and $q_\rmP R$, respectively, where
\begin{align}
q_\rmG &= \frac{M_\rmP}{M_\rmP + M_\rmG(R)} = \frac{q_{\rm enc}(R)}{1+q_{\rm enc}(R)}\,, \nonumber \\
q_\rmP &= \frac{M_\rmG(R)}{M_\rmP + M_\rmG(R)} = \frac{1}{1+q_{\rm enc}(R)}\,.
\label{qgqp}
\end{align}

Throughout this paper, we adopt dimensionless units to describe our dynamical system. All length scales are expressed in units of $r_\rms$, the scale radius of the galaxy, masses are expressed in units of the mass of the galaxy, $M_\rmG$, and velocities are expressed in units of $\sigma = (G M_\rmG / r_\rms)^{1/2}$. The corresponding, characteristic time-scale is $r_\rms/\sigma$.

For convenience, we consider the perturber to be a point mass, but we emphasize that the analysis that follows can be easily extended to accommodate any other (spherically symmetric) perturber potential. In our dimensionless units, we then have that the perturber potential,
\begin{equation}
\Phi_\rmP = -q /r_\rmP.
\label{subjectpotential}
\end{equation}
Throughout we adopt $q = 0.004$ (i.e., the mass of the perturber is only 0.4 percent of that of the galaxy). Unlike the perturbative treatments in TW84 and KS18, though, which require $q$ to be small, our analysis is also valid for more massive perturbers.

\begin{figure}
\includegraphics[width=1\textwidth]{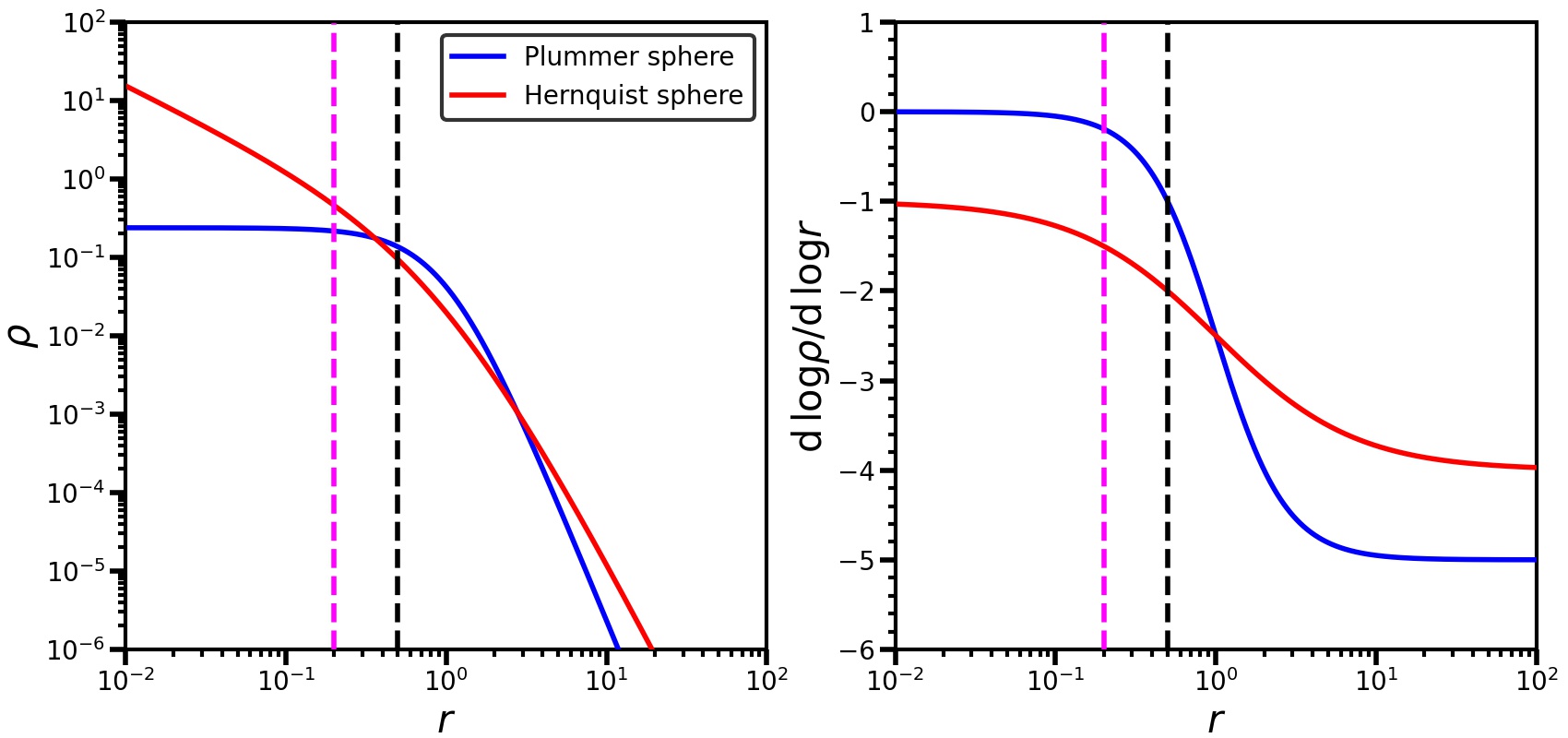}
\caption{\small Density (left-hand panel) and logarithmic slope       $\rmd\log\rho/\rmd\log r$ (right-hand panel) as functions of galacto-centric radius $r$ for the Plummer (blue) and Hernquist (red) spheres used in this paper. The dashed magenta and black lines indicate the orbital radii, $R=0.2$ and $0.5$, considered in this paper. These two radii bracket the bifurcation radius for the Plummer sphere and a $q=0.004$ perturber, at which the torque exerted on the perturber changes from being retarding to enhancing (see sections~\ref{sec:orbits} and \ref{sec:resonances} for details). No such transition occurs for the Hernquist sphere.}
\label{fig:densities}
\end{figure}

In order to contrast dynamical friction in cored and cuspy density profiles, we consider two different density profiles for the galaxy: a Plummer sphere, which has a central constant density core with central logarithmic density gradient, $\gamma \equiv \lim_{r \to 0}\rmd\log\rho/\rmd\log r=0$ \citep{Plummer.11}, and a Hernquist sphere, which has a central $\gamma=-1$ cusp \citep{Hernquist.90}. Both have the advantage that the density and potential are given by simple, analytical expressions. For the Plummer sphere, the density and potential (in our dimensionless units) are given by
\begin{equation}
\rho_\rmG(r) = \frac{3}{4\pi} \frac{1}{\left(1+r^2\right)^{5/2}}\,,
\,\,\,\,\,\,\,\,\,\,\,\,\,
\Phi_\rmG(r) = -\frac{1}{\sqrt{1+r^2}}\,,
\label{plummer}
\end{equation}
while for the Hernquist sphere we have that
\begin{equation}
\rho_\rmG(r) = \frac{1}{2\pi} \, \frac{1}{r \, (1+r)^3}\,,
\,\,\,\,\,\,\,\,\,\,
\Phi_\rmG(r) = -\frac{1}{1+r}\,.
\label{hernquist}
\end{equation}
Figure~\ref{fig:densities} plots these density profiles (left-hand panel) and corresponding logarithmic density gradients, $\rmd\log\rho/\rmd\log r$ (right-hand panel), as functions of radius. The magenta and black vertical dashed lines indicate $R=0.2$ and $0.5$, respectively. These are the orbital radii of the perturber considered in this paper. As we demonstrate below, in the case of the Plummer host these radii bracket the bifurcation radius, $\Rbif$ ($\approx 0.39$ for our fiducial case), at which the orbital make-up of the Plummer sphere undergoes a drastic change due to a bifurcation of some of the Lagrange points, which in turn impacts the nature (retarding vs. enhancing) of the torque on the perturber. In the case of the Hernquist sphere, no such bifurcation occurs.

Throughout, we assume that the galaxies have isotropic velocity distributions, such that their distribution functions are ergodic (i.e., depend only on energy). In the case of the Plummer sphere we have
\begin{align}
f_0(\varepsilon) &= \frac{3}{7\pi^3} {\left(2\varepsilon\right)}^{7/2}\,,
\label{fdistP}
\end{align}
while for the Hernquist sphere 
\begin{align}
f_0(\varepsilon) &= \frac{1}{8 \sqrt{2} \pi^3}\nonumber \\
&\times \frac{3 \sin^{-1}\sqrt{\varepsilon} + \sqrt{\varepsilon (1 - \varepsilon)} (1 - 2 \varepsilon) (8\varepsilon^2 - 8 \varepsilon - 3)}{(1-\varepsilon)^{5/2}}\,.
\label{fdistH}
\end{align}
Here $\varepsilon=-E_{0\rmG}$ ($E_{0\rmG}$ is the unperturbed galactocentric energy), and the subscript `0' indicates that these distribution functions correspond to the unperturbed galaxies. Both distribution functions have been normalized such that 
\begin{align}
\rho_\rmG(r) = 4 \pi \int_0^{\Psi_\rmG} \sqrt{2(\Psi_\rmG-\varepsilon)} \, f_0(\varepsilon) \, \rmd\varepsilon\,,
\end{align}
with $\Psi_\rmG = -\Phi_\rmG$.
\\

\subsection{Hamiltonian dynamics in the co-rotating frame}
\label{sec:Hamilton}

Since the gravitational potential, and hence the Hamiltonian, in the restricted three body problem is time-variable, energy is not a conserved quantity. And due to the lack of spherical symmetry, neither is angular momentum. However, as is well known \citep[see e.g.,][]{Binney.Tremaine.08}, the Jacobi integral,
\begin{equation}
E_{\rm J} = E -{\bf \Omega_{\rm{\bP}}} \cdot \bL = \frac{1}{2}{\dot{\br}}^2 + \Phi_{\rm eff}\left(\br\right),
\label{Ej}
\end{equation}
is a conserved quantity. Here $\br$ is the position vector of the field particle with respect to the COM (see Fig.~\ref{fig:schematic}), and ${\bf \Omega_{\rm{\bP}}} = (0,0,\Omega_\rmP)$ with
\begin{equation}
\Omega_\rmP = \sqrt{\frac{G\,[M_\rmG(R) + M_\rmP]}{R^3}}\,,
\end{equation}
the angular frequency of the perturber with respect to the COM, which, in our dimensionless units, is given by 
\begin{equation}
\Omega_\rmP = \left( \frac{1}{R} \, \left.\frac{\partial \Phi_\rmG}{\partial r_\rmG}\right|_{r_\rmG=R}+\frac{q}{R^3}\right)^{1/2}\,.
\label{Omegap}
\end{equation}
$E$ and $\bL$ are, respectively, the perturbed energy and angular momentum (per unit mass) of the field particle in the non-rotating, inertial frame, given by
\begin{align}\label{Eperturbed}
E& =E_0+\Phi_\rmP=\frac{1}{2}{\vert{\dot{\br}}+\bf{\Omega_{\rm{\bP}}}\times \br \vert}^2 + \Phi_\rmG\left(\br\right) + \Phi_\rmP\left(\br\right),\\
\bL &= \br \times \left(\bf{\dot{r}}+\bf{\Omega_{\rm{\bP}}}\times \br\right)\,.
\end{align}
Here $E_0$ is the unperturbed energy, i.e., the part of the Hamiltonian without the perturber potential, and $\Phi_\rmG$ and $\Phi_\rmP$ are the gravitational potentials due to the galaxy and the perturber, respectively. The effective potential in equation~(\ref{Ej}) is defined as
\begin{equation}
\Phi_{\rm eff}\left(\br\right) = \Phi_\rmG(r_\rmG) + \Phi_\rmP(r_\rmP) - \frac{1}{2} \vert{\bf \Omega_{\rm{\bP}}} \times \br \vert^2,
\label{phieff}
\end{equation}
where $r_\rmG$ and $r_\rmP$ are the distances to the field particle from the galactic center and the perturber respectively, and are given by
\begin{align}
r_\rmG^2 &= r^2 + q_\rmG^2 R^2 + 2\, q_\rmG R\, r \cos{\phi},\nonumber \\
r_\rmP^2 &= r^2 + q_\rmP^2 R^2 - 2\, q_\rmP R\, r \cos{\phi}.
\label{r1r2}
\end{align}
Here $r = \vert \br \vert$, $\phi$ is the counter-clockwise angle between $\br$ and the line connecting the COM and the perturber positioned along the positive $x-$axis (see Fig.~\ref{fig:schematic}), and $q_\rmG$ and $q_\rmP$ are the mass ratios given by equation~(\ref{qgqp}). The third term in equation~(\ref{phieff}) is the potential due to the centrifugal force. Plugging in the expression for $\Omega_\rmP$, and using the fact that $\partial \Phi_\rmG/\partial r = G M_\rmG(r)/r^2$ and $q_{\rm enc}(R)=M_\rmP/M_\rmG(R)$, the effective potential reduces to
\begin{align}
\Phi_{\rm eff}(\br) = \Phi_\rmG(r_\rmG) - \frac{q}{r_\rmP} - \frac{1+q_{\rm enc}(R)}{2}\, \frac{r^2}{R} \, \left.\frac{\partial \Phi_\rmG}{\partial r_\rmG}\right|_{r_\rmG=R}\,.
\end{align}

\begin{table*}
\centering
\tabcolsep=0.03 cm
\begin{tabular}{c|ccc|c|cc}
 \hline
 Orbit type & $E_{\rm{Jc}}^{(4)}$ & $E_{\rm{Jc}}^{(0)}$ & $E_{\rm{Jc}}^\rmP$ & $L$ & COC & Friction (F)/\\
 & & & & & & Buoyancy (B)/ \\
 & & & & & & Negligible (N) \\
 (1) & (2) & (3) & (4) & (5) & (6) & (7) \\
 \hline 
 & & & & & &\\
 \horseshoe $(\gamma=0)$ & $E_{\rm{Jc}}^{(4)}<E_\rmJ^{(3)}$ & $E_{\rm{Jc}}^{(0)}>\max{\left[E_\rmJ^{(1)},E_\rmJ^{(2)}\right]}$  & -- & -- & L3  & F \\ & & & & & & \\
 \horseshoe $(\gamma<0)$ & $E_{\rm{Jc}}^{(4)}<E_\rmJ^{(3)}$ & --  & $E_{\rm Jc}^\rmP>\max{\left[E_\rmJ^{(1)},E_\rmJ^{(2)}\right]}$ & -- & L3  & F \\ & & & & & & \\
 \pacman $(\gamma=0)$ & -- & $E_{\rm{Jc}}^{(0)}<E_\rmJ^{(1)}$ & $E_{\rm{Jc}}^\rmP>E_\rmJ^{(2)}$ & $L^{(1)}<L<L^{(2)}$ & L0 & F/B \\ & & & & & & \\
 \pacman $(\gamma<0)$ & -- & -- & $E_\rmJ^{(2)}<E_{\rm{Jc}}^\rmP<E_\rmJ^{(1)}$ & $L^{(1)}<L<L^{(2)}$ & L0 & F/B \\ & & & & & & \\
 tadpole ($R>\Rbif$) & $E_\rmJ^{(3)}<E_{\rm{Jc}}^{(4)}<E_\rmJ^{(4)}$ & -- & -- & -- & L4/L5 & F/B \\ & & & & & & \\
 tadpole ($R\leq\Rbif$) & $E_\rmJ^{(0)}<E_{\rm{Jc}}^{(4)}<E_\rmJ^{(4)}$ & -- & -- & -- & L4/L5 & F/B \\ & & & & & & \\
\hline
& & & & & & \\
 center-phylic $(\gamma=0)$    & -- &   $E_\rmJ^{(0)}<E_{\rm{Jc}}^{(0)}<E_\rmJ^{(1)}$ & -- & $L<L^{(1)}$ & L0 & N\\ & & & & & & \\
 center-phylic $(\gamma<0)$    & -- &   -- & $E_{\rm Jc}^\rmP<E_\rmJ^{(1)}$ & $L<L^{(1)}$ & L0/cusp & N\\ & & & & & & \\
 perturber-phylic & -- & -- & $E_\rmJ^\rmP<E_{\rm{Jc}}^\rmP<\min{\left[E_\rmJ^{(1)},E_\rmJ^{(2)}\right]}$ & $L^{(1)}<L<L^{(2)}$ & P & N\\ & & & & & & \\
 COM-phylic & -- & -- & $E_{\rm{Jc}}^\rmP<E_\rmJ^{(2)}$ & $L>L^{(2)}$ & COM & N\\ & & & & & & \\
 \hline
\end{tabular}
\caption{\small Different orbital families in the co-rotating frame of our restricted three-body framework. Column (1) indicates the name of the orbital family used throughout this paper. Columns (2), (3) and (4) indicate the bounds on the circular part of the Jacobi energy, $E_{\rm Jc}\approx E_\rmJ-\kappa_0 J_r$ ($\kappa_0$ is the value of the radial epicyclic frequency evaluated at the center of perturbation and $J_r$ is the radial action; see Appendix~\ref{App:orb_class} for details), evaluated in the neighborhood of L4/L5, L0 and the perturber, i.e., $E_{\rm Jc}^{(4)}$, $E_{\rm Jc}^{(0)}$ and $E_{\rm Jc}^\rmP$, respectively. Column (5) indicates the angular momentum, $L$. Column (6) indicates the center-of-circulation (COC), where `P' refers to the perturber, and column (7) indicates whether these orbits contribute significantly to dynamical friction (F) or buoyancy (B) or negligibly to either of the two (N). $E_\rmJ^{(k)}$, with $k=0,1,..,5$, approximately denotes the value of $\Phi_{\rm eff}$ at the $k^{\rm th}$ Lagrange point (see Appendix~\ref{App:orb_class} for details), while $E_\rmJ^\rmP$ denotes that at the location of the perturber ($E_\rmJ^\rmP=-\infty$ for a point mass). $L^{(k)}$, with $k=1,2$, denotes the value of the angular momentum at the $k^{\rm th}$ Lagrange point. Note that \pacman orbits are absent when $E_\rmJ^{(2)}>E_\rmJ^{(1)}$, which is always the case if the galaxy has a central cusp or the perturber is at large $R$ in a cored galaxy. Orbits that are further away from co-rotation resonance can cross the separatrix corresponding to L1, L2 or L3 due to changes in $J_r$, thereby taking on the morphology of a different orbital family, constituting what we call `Chimera orbits' (see section~\ref{sec:orbfam} and Appendix~\ref{App:Chimera} for details).}
\label{tab:Ej}
\end{table*}

\begin{figure*}[t!]
  \centering
  \begin{subfigure}[t]{0.45\textwidth}
    \centering
    \hspace{-2mm}
    \includegraphics[width=.947\textwidth]{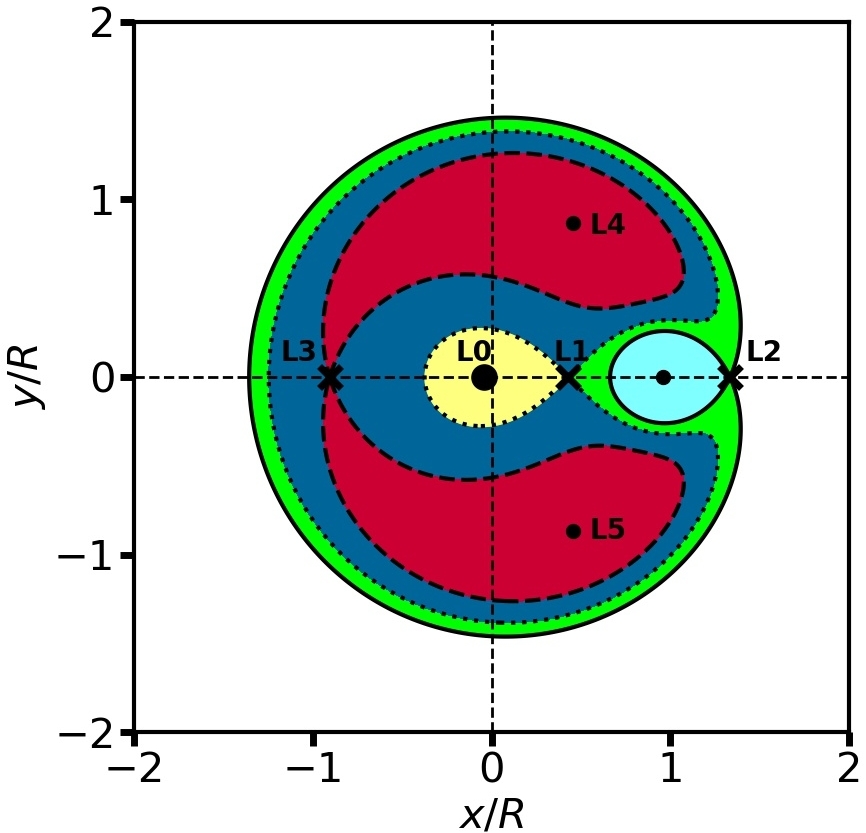}
    \caption{Plummer sphere: outside of the core ($R=0.5$)}
    \label{Plu_out}
  \end{subfigure}
  \begin{subfigure}[t]{0.45\textwidth}
    \centering
    \includegraphics[width=.97\textwidth]{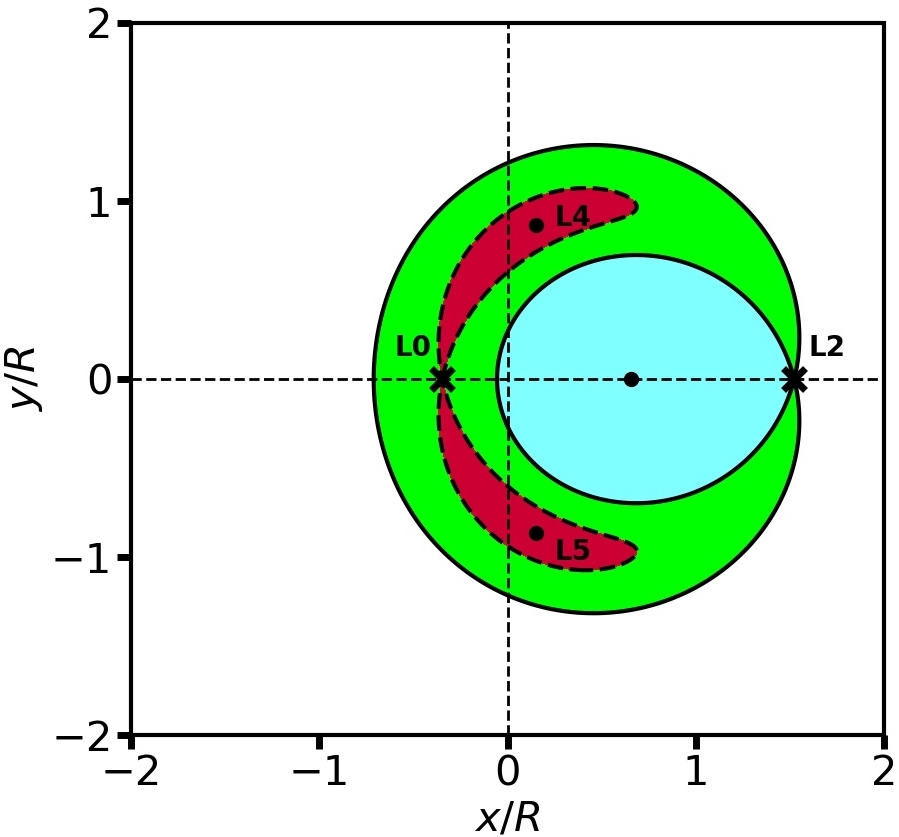}
    \caption{Plummer sphere: inside of the core ($R=0.2$)}
    \label{Plu_in}
  \end{subfigure}
  \\
  \begin{subfigure}[t]{0.45\textwidth}
    \centering
    \includegraphics[width=.945\textwidth]{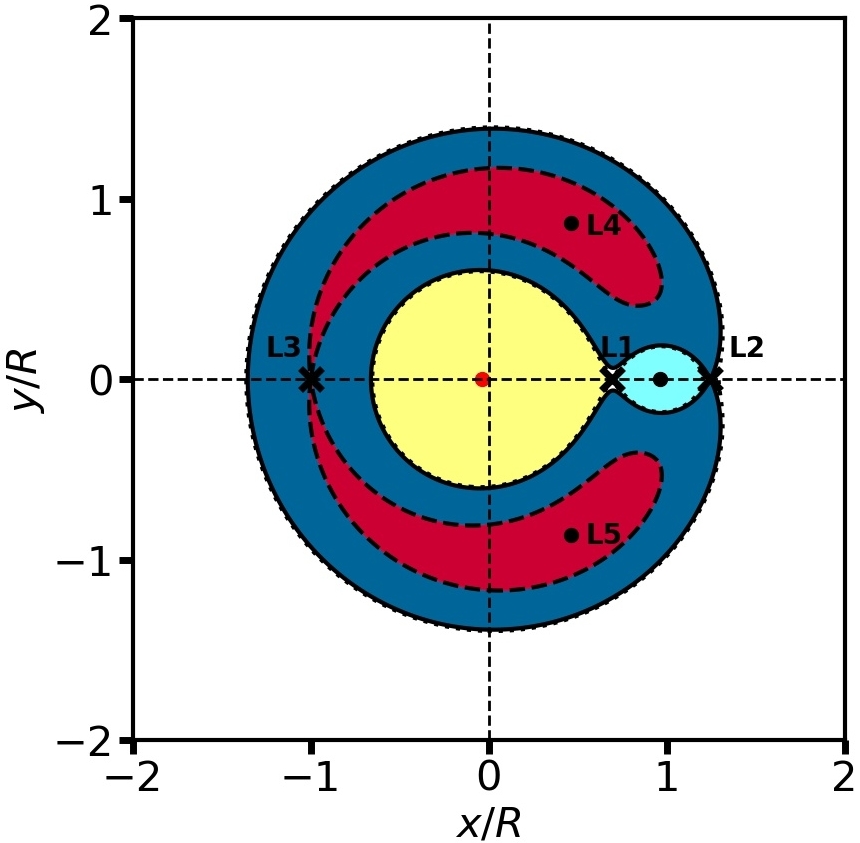}
    \caption{Hernquist sphere ($R=0.5$)}
    \label{Her_out}
  \end{subfigure}
  \begin{subfigure}[t]{0.45\textwidth}
    \centering
    \includegraphics[width=.968\textwidth]{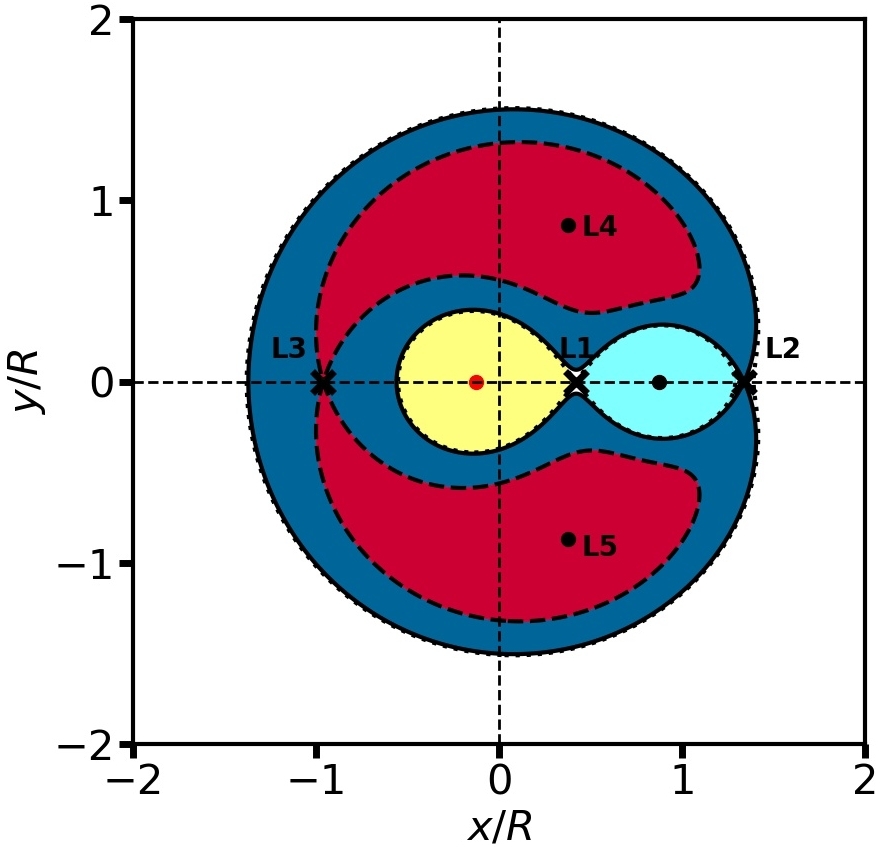}
    \caption{Hernquist sphere ($R=0.2$)}
    \label{Her_in}
  \end{subfigure}
  \caption{\small Effective potential of the galaxy plus perturber with $(x,y)=(0,0)$ corresponding to the COM (see Fig.~\ref{fig:schematic}). The various Lagrange points (fixed points in the co-rotating frame) are indicated, and the different colored regions mark the intervals in Jacobi energy for the zero-velocity curves (ZVCs) of the various near-circular orbital families: \horseshoe (dark blue), \pacman (green), tadpole (red), perturber-phylic (cyan), center-phylic (yellow), and COM-phylic (white). Note that there are no \pacman orbits in a Hernquist galaxy (lower two panels), and that the \horseshoe and center-phylic orbits disappear when the perturber approaches a core (cf. upper two panels). Be aware that the color coding only indicates the locations of the ZVCs: the invariance of the Jacobi energy only limits accessible phase-space from one direction; particles with Jacobi energy $E_\rmJ$ cannot access areas where $\Phi_{\rm eff}(\br) > E_\rmJ$, but given sufficient kinetic energy they can in principle reach any location where  $\Phi_{\rm eff}(\br) < E_\rmJ$. For example, \horseshoe orbits can never enter the red regions, but they can make excursions into the regions that are shaded green, cyan, yellow or white.}
  \label{fig:Phi_eff}
\end{figure*}

\section{Survey of Orbits}
\label{sec:orbits}

\subsection{Equations of motion}
\label{sec:eom}

As already mentioned above, in the perturbed potential, energy and angular momentum are no longer constants of motion. Instead, the only conserved quantity in the restricted three body case considered here is the Jacobi energy,  $E_{\rm J}$. A field particle therefore gains and loses energy and angular momentum (which is exchanged with the perturber) as it traverses its orbit. In order to compute the rates at which the energy and angular momentum of a field particle change as function of time, we integrate its orbit using the equation of motion in the co-rotating frame \citep[][]{Binney.Tremaine.87}, which is given by
\begin{align}
&\ddot{\bf r} = -\nabla \Phi_\rmG - \nabla \Phi_\rmP - 2\left(\bf \Omega_{\rm{\bP}} \times \dot{\bf r}\right) - \bf \Omega_{\rm{\bP}} \times \left(\bf \Omega_{\rm{\bP}} \times \bf r\right)\,.
\label{geneqmotion}
\end{align}
Here the first and second terms on the RHS denote the gravitational accelerations due to the galaxy and the perturber, respectively, while the third and the fourth terms correspond to accelerations due to the Coriolis and centrifugal forces, respectively. In cylindrical coordinates, the above reduces to the following radial and azimuthal equations of motion:
\begin{align}
&\ddot{r} - r\,\dot{\phi}^2 = -\frac{\partial \Phi_\rmG}{\partial r} - \frac{\partial \Phi_\rmP}{\partial r} + 2\,\Omega_\rmP \, r \, \dot{\phi} + \Omega^2_\rmP \, r\,, \nonumber \\
&r \, \ddot{\phi} + 2 \dot{r}\,\dot{\phi} = -\frac{1}{r}\,\frac{\partial \Phi_\rmG}{\partial \phi}-\frac{1}{r}\frac{\partial \Phi_\rmP}{\partial \phi} - 2\,\Omega_\rmP\,\dot{r}\,.
\label{eqmotion}
\end{align}
The latter can be combined with equations~(\ref{r1r2}) to yield an expression for the torque,
\begin{align}
\calT=\frac{\rmd L}{\rmd t}&=-\frac{\partial \Phi_\rmG}{\partial \phi}-\frac{\partial \Phi_\rmP}{\partial \phi}\nonumber \\
&=\frac{r R\, \mathrm{sin}\, \phi}{1+q_{\rm enc}(R)}\left[\frac{q_{\rm enc}(R)}{r_\rmG}\frac{\partial \Phi_\rmG}{\partial r_\rmG}-\frac{1}{r_\rmP}\frac{\partial \Phi_\rmP}{\partial r_\rmP}\right],
\label{singletorque1}
\end{align}
where $L=r^2(\dot{\phi}+\Omega_\rmP)$ is the total angular momentum of the field particle in the inertial frame. Equation~(\ref{singletorque1}) is an expression for the combined torque, exerted by both the perturber and the galaxy on the field particle. For a slowly evolving circular orbit of the perturber, i.e., nearly constant $\Omega_\rmP$, as considered in this paper, $E_\rmJ = E - {\bf \Omega_{\rm{\bP}}} \cdot \bL$ is a conserved quantity. Hence, the corresponding rate of energy change of the field particle is simply given by
\begin{equation}
\frac{\rmd E}{\rmd t} = {\bf \Omega_{\rm{\bP}}} \cdot \frac{\rmd \bL}{\rmd t} \,.
\label{dEdt}
\end{equation}
Because of this equality, throughout this paper we will talk about $\Delta E$ and $\Delta L$ interchangeably. Note that, depending on the sign of the torque $\calT = \rmd L/\rmd t$, the perturber can either lose (dynamical friction) or gain energy (dynamical buoyancy). Also note that dynamical friction or buoyancy results in a non-zero  time-derivative of ${\bf \Omega_{\rm{\bP}}}$, which, following TW84 and KS18, has been ignored in the above equations. Since we are mainly interested in examining dynamical friction near the core-stalling radius, where $\vert \rmd{\bf \Omega_{\rm{\bP}}}/\rmd t \vert$ vanishes, this is justified. In fact, it is justified as long as the time scale for dynamical friction is sufficiently long, i.e., we are in what TW84 refer to as the `slow' regime.

Throughout this paper, all orbit integrations are performed using an exactly Hamiltonian-conserving algorithm proposed by \cite{Kotovych_2002} for simulating general $N-$body systems. It ensures that the Jacobi Hamiltonian is conserved up to machine precision for all the orbits we have integrated. 

\subsection{Orbital Families}
\label{sec:orbfam}

To get a better understanding of dynamical friction, it is instructive to study the different kinds of stellar orbits that arise in presence of the perturber. Using equation~(\ref{geneqmotion}), we numerically integrate stellar orbits in the co-rotating frame under the combined gravitational potential of the perturber plus galaxy. Along each orbit we then register the time-evolution of the orbital energy and angular momentum. We emphasize that in doing so, the perturber is fully accounted for (i.e., is not treated as a small perturbation). For the sake of simplicity, though, we restrict ourselves to 2D, and only study the dynamics in the orbital plane of the perturber. 

One can gain valuable insight regarding the orbital families by examining the system's equipotential contours, which can be parametrized by
\begin{equation}
\Phi_{\rm eff}\left(\bf r\right) = E_{\rmJ}\,.
\end{equation}
These contours are zero-velocity curves (ZVCs) since they map out the locations in the co-rotating frame where the field particles of a given Jacobi energy $E_
\rmJ$ have zero velocity (in the co-rotating frame). Therefore, field particles along an orbit can only occasionally touch its ZVC and can only access regions on the side of its ZVC where its Jacobi energy $E_\rmJ>\Phi_{\rm eff}(\br)$.

Of particular relevance are the fixed points, also known as the Lagrange points, where the effective force in the co-rotating frame vanishes. These are given by the roots of
\begin{equation}
\nabla \Phi_{\mathrm{eff}} = 0\,.
\end{equation}
As we discuss below, and in detail in Paper~II, the number of Lagrange points depends on the inner logarithmic slope $\gamma$ of the galaxy density profile and the galacto-centric distance $R$ of the perturber.

All orbits in the restricted three-body problem have some sense of circulation, either around the galactic center, around the perturber, around the COM, or around a specific Lagrange point.\footnote{The only exceptions are orbits associated with the (stable) Lagrange points, L4 and L5, which are stationary in the co-rotating frame and perfectly circular in the inertial frame.} We can discriminate between these different cases by considering the circular part of their Jacobi energy, $E_{\rm Jc}$, evaluated in the neighborhood of a center of perturbation (COP) (either the location of the perturber or a stable Lagrange point such as L4, L5 or L0),
\begin{align}
E_{\rm Jc}&=E_\rmJ - \frac{{\left(\Delta\Omega_0\right)}^2}{4\left|c_0\right|}\nonumber\\
&- \left(\kappa_0+\frac{b_0}{\left|c_0\right|}\Delta \Omega_0\right)J_r - \left(a_0+\frac{b^2_0}{\left|c_0\right|}\right)J^2_r.
\label{Ejcirc}
\end{align}
Here $J_r$ is the radial action, $\Delta\Omega_0=\Omega_0-\Omega_\rmP$, $\Omega_0$ and $\kappa_0$ are the azimuthal and radial epicyclic frequencies, respectively, and $a_0$, $b_0$ and $c_0$ are constants that depend on the galaxy potential, evaluated at the COP (see Appendix~\ref{App:orb_class} for details). The family of an orbit is dictated by the values of $E_{\rm Jc}$ computed in the neighborhood of L4/L5 ($E_{\rm Jc}^{(4)}$), L0 ($E_{\rm Jc}^{(0)}$) and the perturber ($E_{\rm Jc}^\rmP$) respectively, relative to the values of the effective potential, $\Phi_{\rm eff}$, at the various Lagrange points and the location of the perturber. In what follows, we use $E_\rmJ^{(k)}$, with $k=0,1,..,5$ to (approximately) indicate the value of $\Phi_{\rm eff}$ at the $k^{\rm th}$ Lagrange point (e.g., $E_\rmJ^{(3)}$ indicates the $\Phi_{\rm eff}$ value corresponding to the equipotential/zero-velocity contour that passes through L3), and $E_\rmJ^\rmP$ to indicate the value of the effective potential at the location of the perturber (see Appendix~\ref{App:orb_class} for details). For nearly circular orbits with $J_r\approx 0$, $E_{\rm Jc}\approx E_\rmJ$ and the orbital families are roughly dictated by the equipotential contours.

\begin{figure*}[t!]
  \centering
  \begin{subfigure}{1.03\textwidth}
    \centering
    \includegraphics[width=0.93\textwidth]{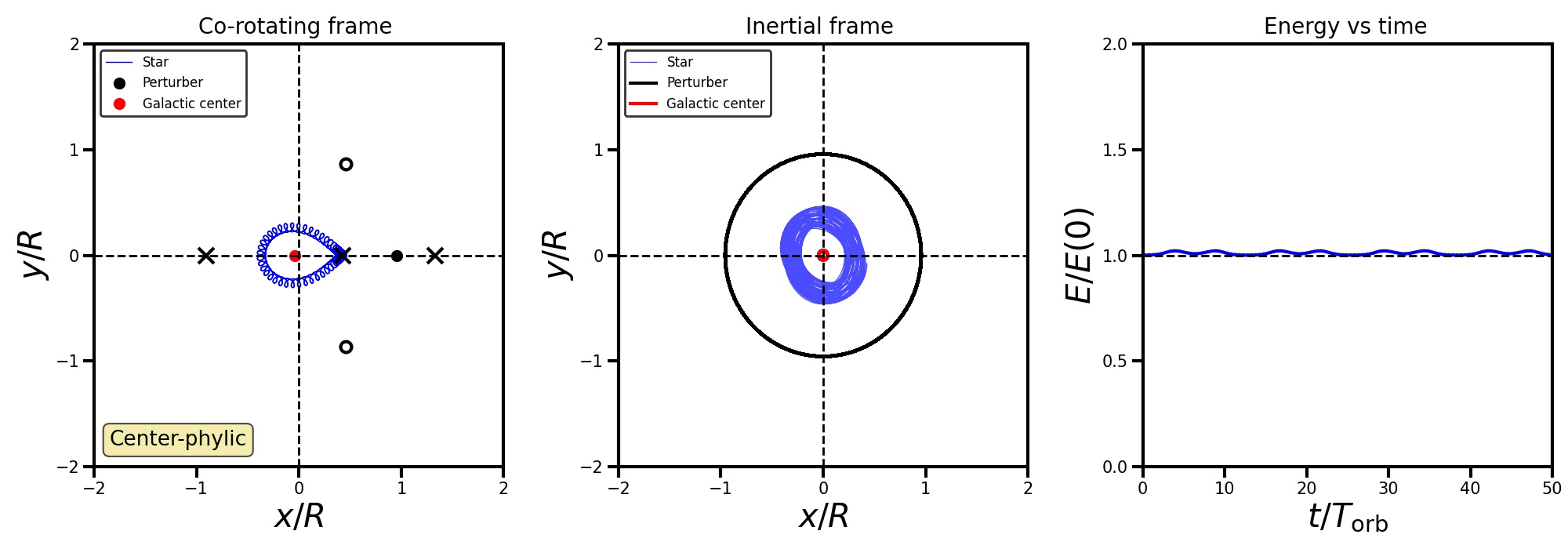}
    \label{orbitphyla}
  \end{subfigure}
  \\
  \begin{subfigure}{1.03\textwidth}
    \centering
    \includegraphics[width=0.93\textwidth]{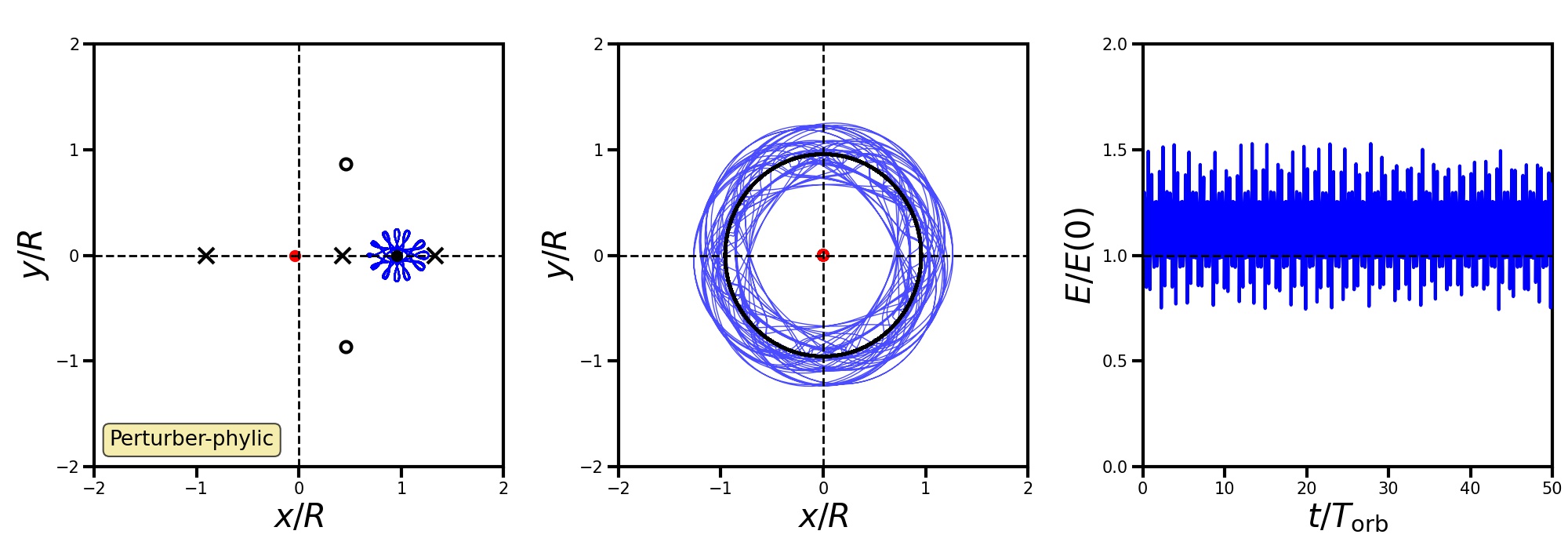}
    \label{orbitphylb}
  \end{subfigure}
  \\
  \begin{subfigure}{1.03\textwidth}
    \centering
    \includegraphics[width=0.93\textwidth]{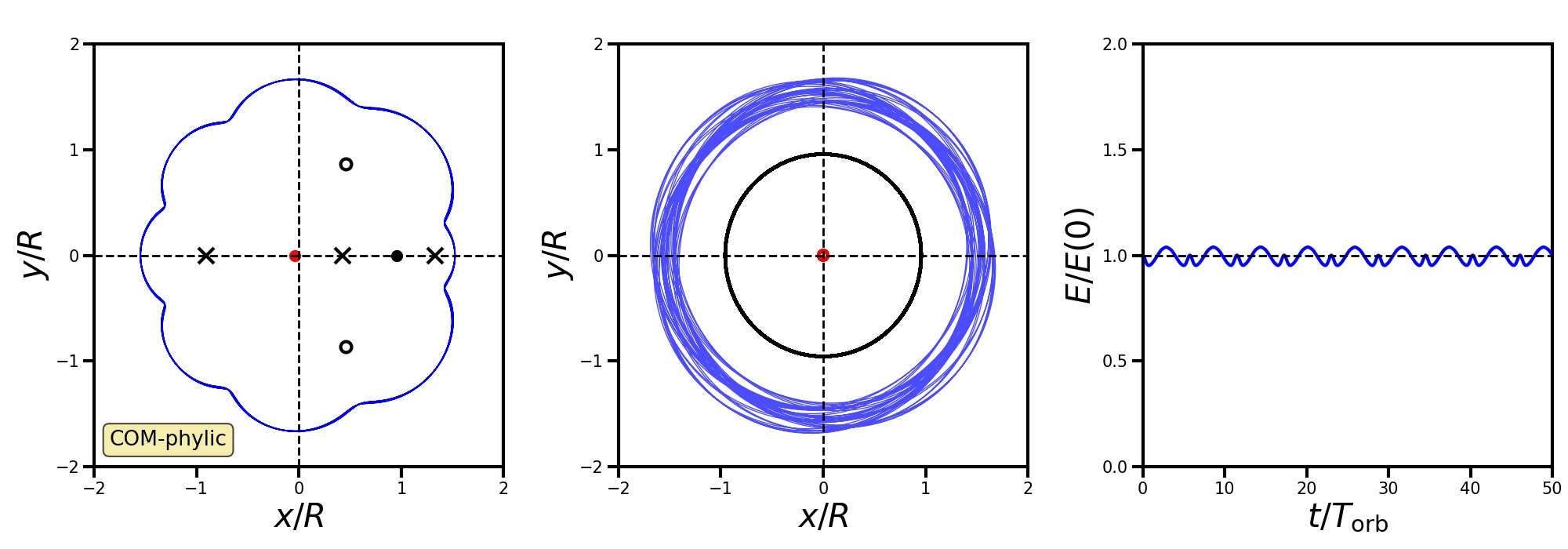}
    \label{orbitphylc}
  \end{subfigure}
  \caption{\small Three orbital families (from top to bottom, center-phylic, perturber-phylic and COM-phylic) in a Plummer sphere with a perturber ($q=0.004$) on a circular orbit outside the core ($R=0.5$). As always, (x,y)=(0,0) corresponds to the COM (see Fig.~\ref{fig:schematic}). In each row, the left-hand panel shows the orbit in the co-rotating frame. The black dot indicates the perturber, the red dot marks the galactic center, and the open circles and crosses mark the stable and unstable Lagrange points, respectively. The middle panels show the orbits in the inertial frame, and the right-hand panels show the evolution in energy (as a function of time in units of $T_{\rm orb}$, the orbital time of the perturber) for a particle moving along the orbit. As discussed in the text, none of these orbital families significantly contribute to dynamical friction.}
  \label{fig:orbit1}
\end{figure*}

We start our census of the orbital families by considering a Plummer galaxy with a massive perturber (as always, assumed to be a point mass with $q=0.004$) orbiting at $R=0.5$, which is outside of the bifurcation radius (see Fig.~\ref{fig:densities}). The equipotential contours ($\Phi_{\rm eff} = E_\rmJ$) in this case are as depicted in Fig.~\ref{Plu_out}. The system has 6 Lagrange points 
(L0, L1, L2 , L3, L4 and L5) as indicated. Of these, L0 (which coincides with the galactic center), L1, L2 and L3 all lie along the $x$-axis, while L4 and L5 are located symmetrically on both sides of it, each forming an equilateral triangle with L0 and the perturber. As discussed in detail in Paper~II, the Lagrange points L0, L4 and L5 are stable fixed points (centers), while L1, L2 and L3 are unstable fixed points (saddles).

We identify different orbital families based on the circular part of the Jacobi energy, $E_{\rm Jc}$, as specified in Table~\ref{tab:Ej}; for nearly circular orbits, this amounts to considering only the Jacobi energy since they lie close to their ZVCs. All such near-circular orbits with ZVCs inside the same shaded region in Fig.~\ref{Plu_out} have similar morphology and are taken to belong to the same orbital family. These families are separated by the ZVCs passing through the saddle points, known as separatrices. Note though that since $J_r$ can vary along an orbit, certain orbits (especially those with higher eccentricities in the inertial frame) can transition between different orbital families by undergoing separatrix-crossings. We shall address these special kinds of orbits separately towards the end of this section and proceed with the delineation of orbital families using $E_{\rm Jc}$ for now.

Let's start with the yellow-shaded region in Fig.~\ref{Plu_out}. These are orbits that circulate the galactic center (which coincides with the stable Lagrange point L0 for $\gamma>-1$ and the central cusp for $\gamma\leq -1$). These are characterized by $E_\rmJ^{(0)} < E_{\rm Jc}^{(0)} < E_\rmJ^{(1)}$ for central cores ($\gamma=0$) and $E_{\rm Jc}^\rmP<E_\rmJ^{(1)}$ for steeper profiles ($\gamma<0$). Additionally, they have lower angular momentum than that at L1, $L^{(1)}$, i.e., have $L<L^{(1)}$. Their orbital frequency is typically much larger than that of the perturber, and particles on these orbits are thus far from co-rotation resonance. In what follows we shall refer to such orbits as `center-phylic'. An example is shown in the top row of Fig~\ref{fig:orbit1}. As is evident from the right-hand panel, the orbital energy varies very little with orbital phase. As a consequence, field particles on these center-phylic orbits exchange very little energy with the perturber, and thus do not contribute significantly to dynamical friction. 

There is a similar family of non-resonant orbits, with $E_\rmJ^\rmP < E_{\rm Jc}^\rmP < \min[E_\rmJ^{(1)},E_\rmJ^{(2)}]$, that, in the co-rotating frame, only circulate the perturber. These orbits, which we call `perturber-phylic', are restricted to the Roche-lobe centered on the perturber (shaded light-blue in Fig.~\ref{Plu_out}). Their angular momentum is higher than that at L1, $L^{(1)}$, but smaller than that at L2, $L^{(2)}$, i.e., they have $L^{(1)}<L<L^{(2)}$. An example is shown in the middle row of Fig~\ref{fig:orbit1}. Note that, due to the proximity to the perturber, the orbital energy along this orbit changes drastically, and rapidly. Because of the rapid oscillations of orbital energy, the {\it net} energy exchange from {\it all} field particles on these perturber-phylic orbits is negligible, and this orbital family therefore is also not a significant contributor to dynamical friction.

Next, there is a family of low-$E_\rmJ$ orbits with $E_{\rm Jc}^\rmP < E_\rmJ^{(2)}$, that circulate the COM of the combined galaxy$+$perturber system. Their ZVCs (for near-circular orbits) fall in the unshaded region of Fig.~\ref{Plu_out} (outside of the equipotential contour that passes through L2), as their angular momentum prevents them from entering the `central' (shaded) regions, i.e., they have $L>L^{(2)}$. An example of such a `COM-phylic' orbit is shown in the bottom row of Fig~\ref{fig:orbit1}. It reveals small fluctuations in orbital energy on a relatively short timescale. Since there are roughly equal numbers of field particles along each phase of these COM-phylic orbits, they also have a negligible, net contribution to dynamical friction (i.e., at each point in time, these orbits contribute roughly equal numbers of energy gainers as energy losers).

\begin{figure*}[t!]
  \centering
  \begin{subfigure}{1.03\textwidth}
    \centering
    \includegraphics[width=0.93\textwidth]{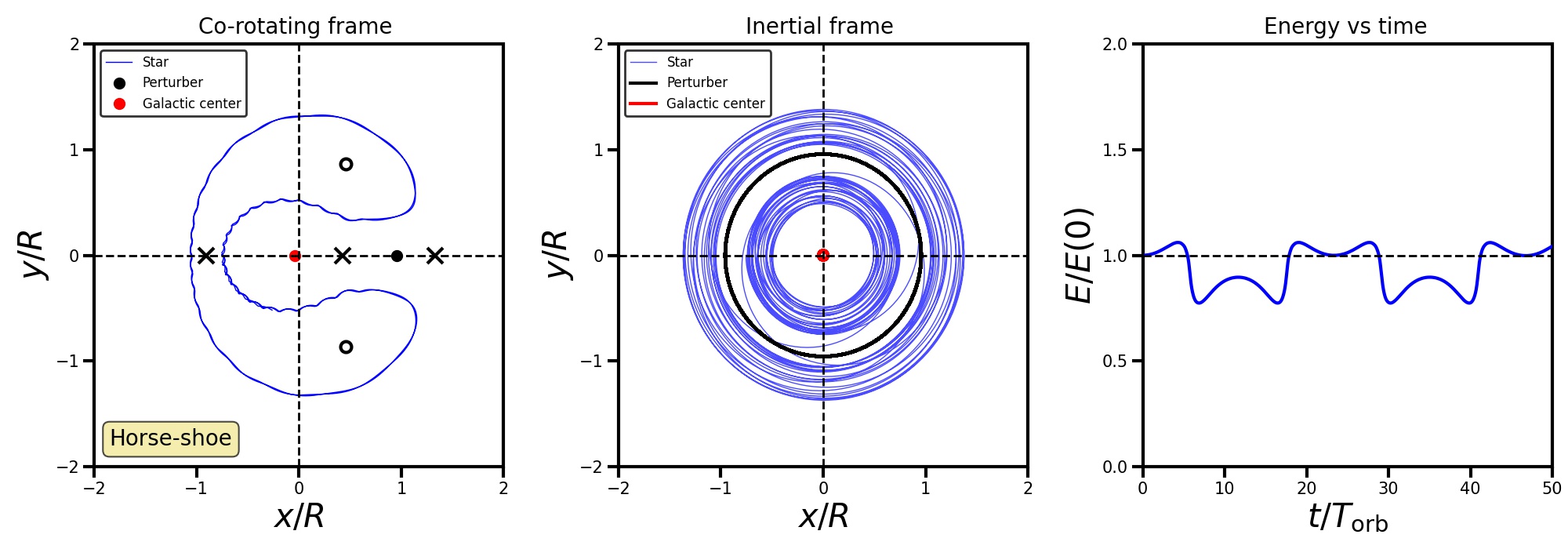}
    \label{orbitouta}
  \end{subfigure}
  \\
  \begin{subfigure}{1.03\textwidth}
    \centering
    \includegraphics[width=0.93\textwidth]{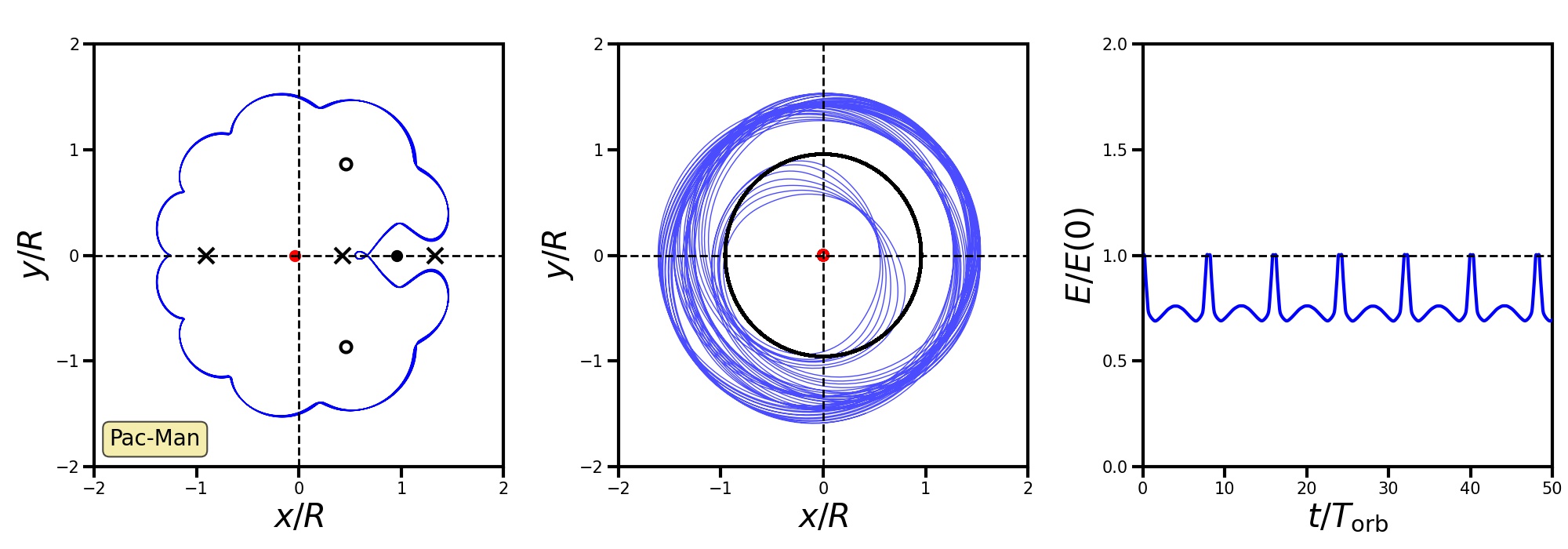}
    \label{orbitoutb}
  \end{subfigure}
  \\
  \begin{subfigure}{1.03\textwidth}
    \centering
    \includegraphics[width=0.93\textwidth]{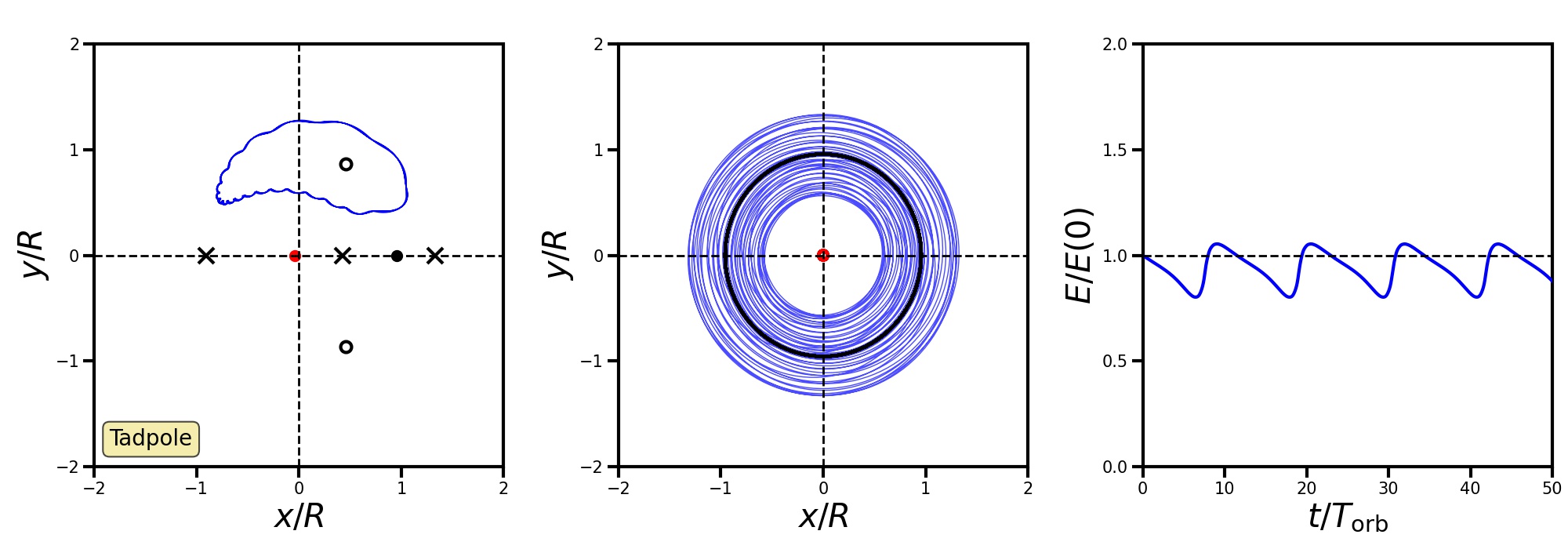}
    \label{orbitoutc}
  \end{subfigure}
  \caption{\small Same as Fig~\ref{fig:orbit1}, but for the three NCRR families (from top to bottom, \horseshoen, \pacman and tadpole) that make significant contribution to dynamical friction.}
  \label{fig:orbit2}
\end{figure*}

Next, we discuss the three families that are the dominant contributors to dynamical friction. They all have azimuthal frequencies that are comparable to that of the perturber, i.e., $\Omega_\phi\approx \Omega_\rmP$, such that their libration time in the co-rotating frame is long. In fact, along these orbits, $\Omega_\phi-\Omega_\rmP$ oscillates back and forth about $\pm\Omega_r/N$, where $\Omega_r$ is the radial frequency and the integer $N$ is the number of radial excursions or epicycles for every libration. The typical range of $N$ is $[2,\infty)$ for realistic galaxy profiles, with $N\to \infty$ marking the co-rotation resonance, i.e., $N$ is larger the closer the orbit is to co-rotation. Therefore they are `near-co-rotation-resonant' (NCRR), i.e., they librate about the near-co-rotation resonances, $\Omega_\phi-\Omega_\rmP=\pm\Omega_r/N$. When the perturber is farther out, $M_\rmG(R)\gg M_\rmP$, implying $\Omega_\rmP\approx \sqrt{GM_\rmG(R)/R^3}\approx \Omega_\phi$ in the vicinity of L4 and L5 (since these two Lagrange points are both at a distance, $R$, from the galactic center). Therefore, $N$ is large, i.e., $N\gg 1$, and the orbits librating about L4 and L5 are close to co-rotation resonance. As the perturber penetrates deeper into the core region, $M_\rmP$ becomes comparable to $M_\rmG(R)$, and $\Omega_\rmP$ significantly exceeds $\Omega_\phi$ near L4 and L5, thereby pushing the orbits farther away from co-rotation resonance (smaller $N$), as pointed out by KS18.

The first, and probably most well-known, among the NCRR orbits is the family of so-called `\horseshoe orbits', which we already encountered in Section~\ref{sec:concept}. These have $E_{\rm Jc}^{(4)} < E_\rmJ^{(3)}$ and $E_{\rm Jc}^{(0)} > \max[E_\rmJ^{(1)},E_\rmJ^{(2)}]$ for central cores ($\gamma=0$), while $E_{\rm Jc}^{(4)} < E_\rmJ^{(3)}$ and $E_{\rm Jc}^\rmP>\max[E_\rmJ^{(1)},E_\rmJ^{(2)}]$ for steeper profiles ($\gamma<0$). The ZVCs of near-circular \horseshoe orbits fall within the dark blue-shaded region in Fig.~\ref{Plu_out} and can only cross the $x$-axis at the side of L0 opposite to the perturber; the Lagrange point L1 acts as a barrier, forcing the particle to take a long `detour' around the center of the galaxy. They have a net sense of circulation around L3, with a libration frequency $\vert \Omega_{\rm lib} \vert \ll \Omega_\rmP$. As is evident from the top row of Fig.~\ref{fig:orbit2} (see also Fig.~\ref{fig:horseshoe}), the orbital energy can vary drastically along the orbit, undergoing rapid changes when close to the perturber, where the perturber's force pulls the field particle either inward or outward. 

Somewhat similar to the \horseshoe orbits is a family of orbits that we call `\pacmann' orbits. These are characterized by $E_{\rm Jc}^{(0)} < E_\rmJ^{(1)}$ and $E_{\rm Jc}^\rmP>E_\rmJ^{(2)}$ for central cores ($\gamma=0$), while $E_\rmJ^{(2)}<E_{\rm Jc}^\rmP<E_\rmJ^{(1)}$ for steeper profiles ($\gamma<0$). Additionally, they have $L^{(1)}<L<L^{(2)}$. They differ from the \horseshoe orbits in that they have a net sense of circulation around L0. The Jacobi energy of the near-circular \pacman orbits is less than that of L1, which allows their ZVCs to cross the $x$-axis at the side of L0 that coincides with the perturber, and fall within the green-shaded region of Fig.~\ref{Plu_out}. Rather than taking a `detour', these orbits can therefore take a `short-cut', which changes their characteristic shape such that they resemble the iconic flashing-dots eating character of the popular 1980's computer-game Pac-Man (see middle row of Fig.~\ref{fig:orbit2}). We emphasize that \pacman orbits are only present when $E_\rmJ^{(1)} > E_\rmJ^{(2)}$. For a given galaxy potential and mass of the perturber, this puts a constraint on the galacto-centric distance of the perturber, $R$; for the Plummer potential and our fiducial mass ratio $q=0.004$, \pacman orbits are only present when the perturber is located at $R \lta 1.23$. When further out, \pacman orbits are absent such that the equipotential contours and orbital families are similar for both cored and cuspy galaxy profiles. 

The final family of NCRR orbits are known as `tadpole' orbits, a name that again relates to their characteristic shape in the co-rotating frame (see bottom row of Fig.~\ref{fig:orbit2}). These are characterized by $E_\rmJ^{(3)} (E_\rmJ^{(0)}) < E_{\rm Jc}^{(4)} < E_\rmJ^{(4)}=E_\rmJ^{(5)}$ for $R>\Rbif$ ($R\leq \Rbif$), and have a net sense of circulation around either L4 or L5. Their ZVCs fall within the red-shaded region of Fig.~\ref{Plu_out}.

\subsection{Slow versus fast actions}
\label{sec:slowfast}

Along all NCRR orbits (\horseshoen, \pacman and tadpole), the energy and angular momentum oscillate with a large amplitude and long period, and the star is up/down-scattered through near-co-rotation resonances by interactions with the perturber. This can be understood in terms of slow and fast action-angle variables, which exist in the neighborhood of a resonance and are related to the radial and azimuthal action-angle variables by a canonical transformation \citep[e.g.,][]{Tremaine.Weinberg.84,Lichtenberg.Lieberman.92,Chiba.Schonrich.21}. The NCRR orbits librate about the commensurability condition $\Omega_\phi -\Omega_\rmP \mp \Omega_r/N = 0$. The corresponding angle, $\theta_s = \theta_\phi \mp \theta_r/N - \Omega_\rmP t$, is called the {\it slow angle}, and the action conjugate to it is called the slow action, $J_s$, which is proportional to the angular momentum. Note that close to the commensurability condition $\rmd\theta_s/\rmd t = \Omega_\phi - \Omega_\rmP \mp \Omega_r/N \simeq 0$, indicating that $\theta_s$ indeed varies slowly. And while it does, the corresponding slow action undergoes large changes.  Both $J_s$ and $\theta_s$ librate about the near-co-rotation resonances with a time period, $T_{\rm lib}$, which is much larger than the orbital time of the perturber \citep[see][for detailed derivations using perturbative expansions of the Hamiltonian around resonances]{Contopoulos.73, Chiba.Schonrich.21}. In fact, for orbits that come arbitrarily close to the separatrices, $T_{\rm lib}$ approaches infinity.

Contrary to the slow angle, the {\it fast angle}, which is nothing but the radial angle, $\theta_r$, varies rapidly along an orbit, while its conjugate action, the {\it fast action}, $J_f=J_r\pm L/N$, is nearly invariant. In general, the faster the angle changes, the closer its corresponding fast action is to an adiabatic invariant. Therefore, the NCRR orbits have two integrals of motion, the Jacobi Hamiltonian, $E_\rmJ$ (which is exactly conserved), and the fast action, $J_f$ (which is {\it very nearly} conserved), and are {\it nearly integrable}\footnote{In 3D, the near-resonant orbits possess a second pair of fast action-angle variables, where the fast angle corresponds to the azimuthal angle along the orbital plane of the field particle, which can be inclined wrt the perturber's plane of orbit.}. For the very nearly co-rotation resonant orbits, $N\gg 1$, and therefore $J_f\approx J_r$, i.e., the orbital eccentricity (in the inertial frame) remains nearly constant. This is however not the case for orbits farther away from co-rotation resonance, which can show very interesting dynamics, as we shall see shortly.

\subsection{Orbital make-up}
\label{sec:makeup}

The relative abundances of the different orbital families depend on the orbital radius $R$ of the perturber. For example, Fig.~\ref{Plu_in} shows the equipotential contours of the same Plummer galaxy as in Fig.~\ref{Plu_out}, but with the perturber orbiting inside the central core, at $R = 0.2$. Now only four Lagrange points are present; both L1 and L3 have disappeared. As the perturber approaches the galactic center, the Roche lobes around the galactic center and the perturber coalesce to form a single lobe surrounding the perturber. As we show in Paper~II, this is associated with the merging, or `bifurcation' of L3, L0 and L1 at a critical bifurcation radius, $\Rbif$, which leaves only L0, L2, L4 and L5, and changes the stability of L0 from being a center to a saddle.  As a consequence, neither \horseshoe nor center-phylic orbits survive. In addition, the contribution of the tadpole orbits is also significantly diminished. Instead, the dominant orbital families in the central core region are the perturber-phylic orbits and the \pacman orbits. As we will see, this has profound implications for dynamical friction. 

The orbital configuration is particularly sensitive to the density profile of the galaxy. The lower two panels of Fig.~\ref{fig:Phi_eff} show the equipotential contours of a Hernquist galaxy with a perturber at $R=0.5$ (Fig.~\ref{Her_out}) and $R=0.2$ (Fig.~\ref{Her_in}). In such a cuspy galaxy, there is no L0 (L0 is replaced by the cusp), and the five Lagrange points (L1, L2, L3, L4 and L5) survive throughout, for any value of the orbital radius of the perturber, $R$, without the occurrence of any bifurcation. As a consequence, in this galaxy potential, there are never any \pacman orbits and the relative abundances of different orbital families show a much weaker dependence on $R$ than in the case of the Plummer sphere. How all of this relates to dynamical friction will be discussed in more detail in sections~\ref{sec:resonances}-\ref{sec:core}.

\subsection{Separatrix crossing and Chimera orbits}
\label{sec:crossing}

Before proceeding with the computation of the dynamical friction torque from the various orbits, we first discuss a potential complication. We have defined orbital families on the basis of $E_{\rm Jc}$, but family is not an invariant property for all orbits. In fact, an orbit can change its family in course of its evolution. This is because the orbit-determinant, $E_{\rm Jc}$, as expressed in equation~(\ref{Ejcirc}), is not an invariant quantity. It not only involves $E_\rmJ$, which is an integral of motion and thus conserved, but also the radial action, $J_r$, which is typically not constant along an orbit. In particular, $J_r$ can undergo significant changes along orbits that are farther away from co-rotation resonance, since only a linear combination of $J_r$ and $L$, and not $J_r$ alone, is the fast action in this case. Therefore the value of $E_{\rm Jc}$ can potentially cross over from that corresponding to one orbital family to another, which corresponds to the orbit undergoing separatrix-crossing due to a change in the radial action enabled by the perturber, altering its morphological appearance. We call such orbits `Chimera orbits'\footnote{The Chimera orbits are named after the hybrid creature in Greek mythology that is composed of parts of more than one animal.}. These Chimera-like transitions occur between trapped regions of neighboring resonances on either side of a separatrix (see Appendix~\ref{App:orb_class}) or a chaotic island formed by the {\it overlap} of resonances \citep[see][for a detailed discussion in the context of bar-like perturbations]{Chiba.Schonrich.21}. For example, the metamorphosis between \horseshoes and tadpoles occurs near L3, while that between \horseshoens, \pacmans and center-phylic orbits happens near L1. And finally the transition between \pacmann, COM-phylic and perturber-phylic orbits occurs in the neighborhood of L2. We show several examples of such Chimera orbits in Appendix~\ref{App:Chimera}. Not all orbits show this Chimera behavior. The very nearly co-rotation resonant orbits are nearly circular and thus have small $J_r$. Since $J_r$ is a fast action along such orbits, it remains almost constant, i.e., the orbits remain nearly circular and do not exhibit Chimera characteristics. 

When the separatrix crossing along a Chimera orbit results in a perturber-phylic phase, we speak of resonant capture \citep[][]{Henrard.82}, which as pointed out in \cite{Tremaine.Weinberg.84}, can `dress' the perturber with a cloud of captured stars. Note, though, that in the `slow' regime considered here, in which the orbital radius of the perturber is taken to be invariant, these stars can undergo separatrix crossing again, transitioning back to a \pacman or a COM-phylic orbit. Similarly, when a separatrix-crossing results in a transition from a `trapped' NCRR state to an `untrapped' COM-phylic state, the transition is sometimes called `scattering', e.g., \cite{Daniel.Wyse.15}.

Chimera orbits are difficult to account for in our treatment because they do not have a clear periodic behaviour, i.e., do not have a well-defined libration time. However, we find that most of them typically behave as an archetypal orbit of their family for many orbital periods before revealing their Chimera nature, i.e., they are `semi-ergodic' (similar to the semi-ergodic orbits identified by \cite{Athanassoula.etal.83} in their study of barred galaxies). This is akin to how Arnold diffusion in KAM theory can cause chaotic orbits to behave quasi-regularly for extended periods  \citep[e.g.][]{Lichtenberg.Lieberman.92}. Hence, we conjecture that their relevance to dynamical friction is captured, at least to leading order, by our following treatment of the NCRR orbital families.

\begin{figure}
\centering
\hspace{-1mm}
\includegraphics[width=1\textwidth]{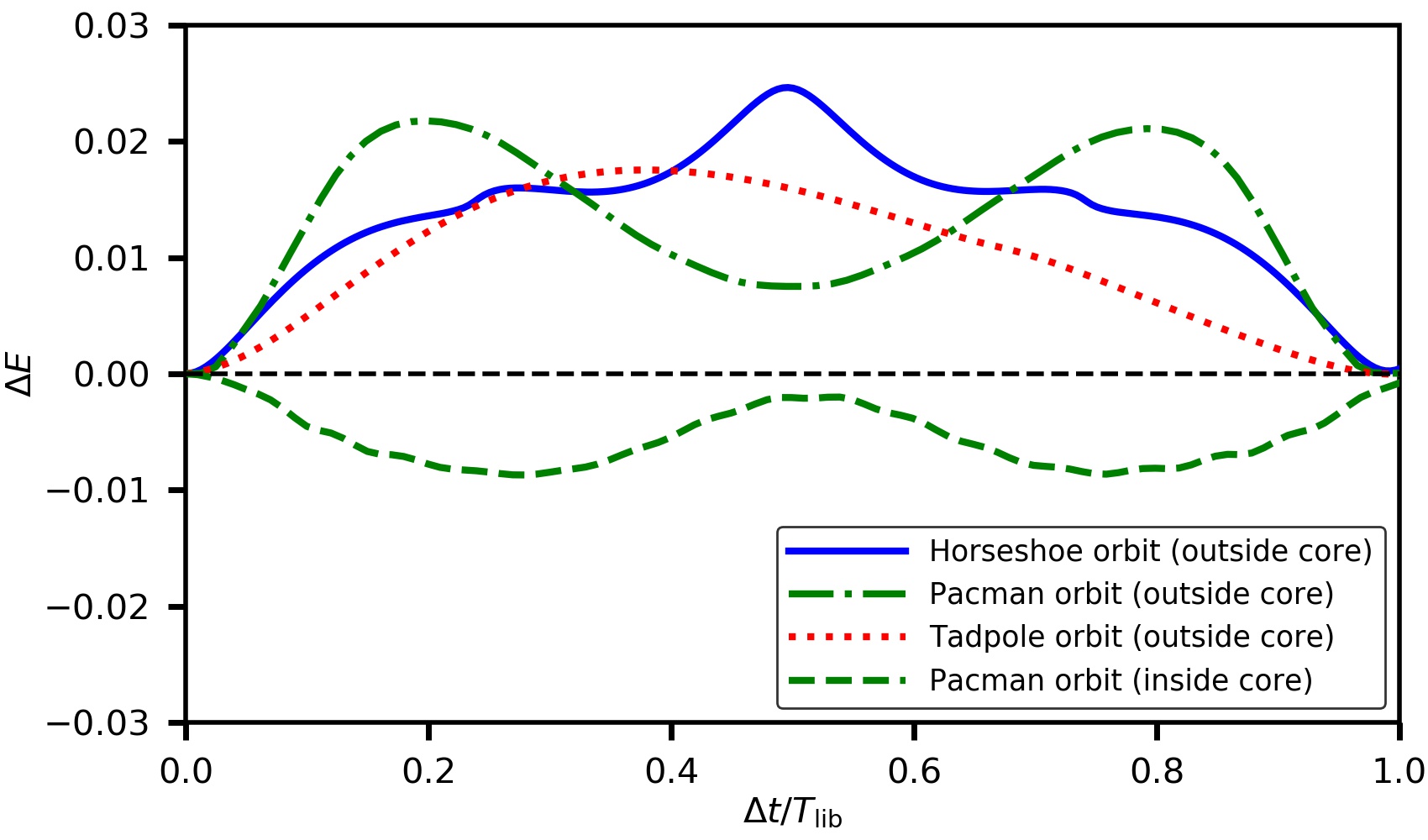}
\caption{\small The solid blue, dot-dashed green and dotted red curves respectively show the average energy change per star (equation~[\ref{deltaE}]) along individual NCRR \horseshoen, \pacman and tadpole orbits shown in Fig.~\ref{fig:orbit2} as a function of time (in units of the libration time, $T_{\rm lib}$). All these are examples of orbits in the case where the perturber is orbiting outside of the core of a Plummer sphere, at $R=0.5$. For comparison, the green dashed curve shows the integrated energy change for a \pacman orbit when the perturber is orbiting inside the core, at $R=0.2$. See text for details.}
\label{fig:integrated_energy}
\end{figure}

\section{The origin of dynamical friction in the non-perturbative case} 
\label{sec:resonances}

As described in Section~\ref{sec:concept}, in our non-perturbative framework the net torque on the perturber arises from an {\it imbalance} between field particles {\it along the same orbit} that are {\it up-scattered} vs. {\it down-scattered} in energy. We now proceed to compute the torque on the perturber due to individual orbits. Using the results from a large ensemble of such orbits, we then highlight the transition from a net retarding to a net enhancing torque when approaching the core of a Plummer sphere.

\begin{figure*}[t!]
  \centering
  \begin{subfigure}[t]{0.48\textwidth}
    \centering
    \includegraphics[width=1\textwidth]{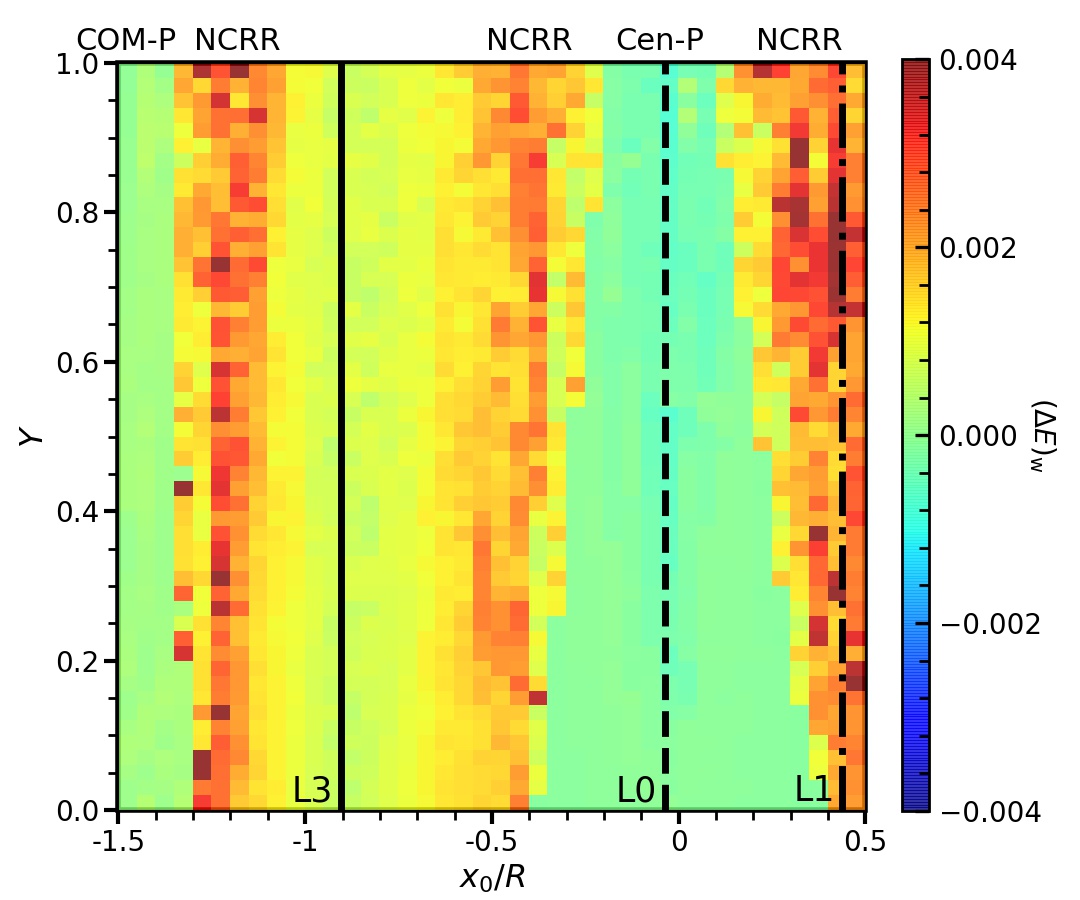}
  \end{subfigure}
  \hspace{2mm}
  \begin{subfigure}[t]{0.49\textwidth}
    \centering
    \includegraphics[width=1\textwidth]{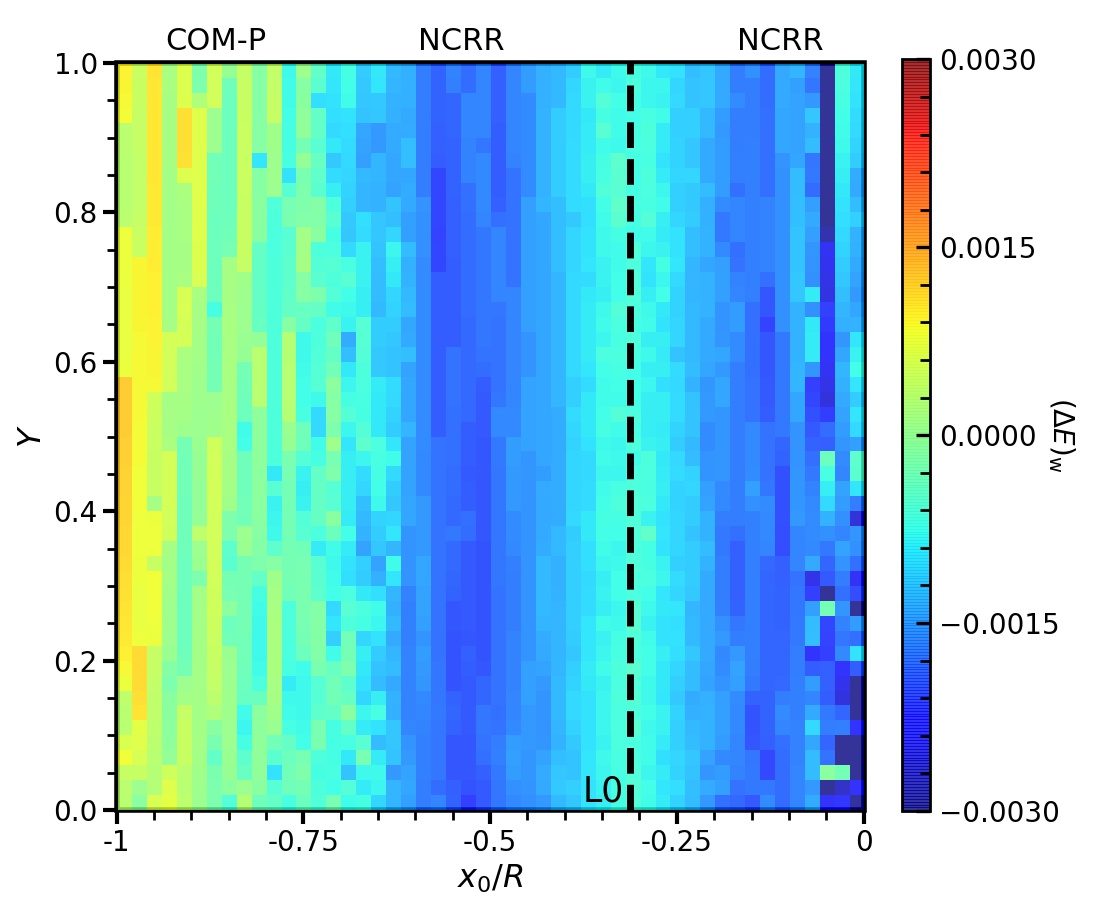}
  \end{subfigure}
  \caption{\small Energy change per unit phase space, $\dE$, of field particles moving along orbits in a cored Plummer potential with a perturber $(q=0.004)$ on a circular orbit at $R=0.5$ (left-hand panel) and $R=0.2$ (right-hand panel). The initial conditions for the orbits are sampled uniformly in $x_0$ and $Y \equiv [E_\rmJ  -\Phi_{\rm eff} (x_0,0)] / [E_\rmJ^{(4)} - \Phi_{\rm eff}(x_0,0)]$ (for every $x_0$), with $y_0=0$ and $\left|v_{\rmy,0}\right| = \frac{1}{3}v_0$, where $v_0 = \sqrt{2[E_\rmJ-\Phi_{\rm eff}(x_0,0)]}$.  Solid, dashed, and dot-dashed vertical lines indicate the positions of L3, L0 (the galactic center) and L1, respectively. Note that when the perturber is located outside the core, at $R=0.5$, $\dE$ is predominantly positive (red) suggesting ongoing dynamical friction. Inside the core, though, at $R=0.2$, $\dE$ is predominantly negative (blue) indicating dynamical buoyancy. The red and blue bands are due to NCRR orbits (causing a larger $|\dE|$), while bands of greenish color (small $|\dE|$) generally indicate non-resonant orbits. In particular, the wide green band in the left panel centered on $x_0=0$ corresponds to the non-resonant center-phylic (Cen-P) orbits, while the green band in the extreme left of both panels indicates COM-phylic (COM-P) orbits. As discussed in the text, due to a bifurcation of Lagrange points there are no center-phylic orbits when the perturber is inside $R \sim 0.39$.}
  \label{fig:delE}
\end{figure*}
\subsection{The net torque from individual orbits}
\label{sec:integrated_energy}

In order to compute the torque on the perturber due to a single orbit, we proceed as follows. We numerically integrate the orbit of a massless field particle in the presence of the perturber, registering its position $\br$, velocity $\dot{\br}$, energy $E$, and angular momentum $\bL$, as a function of time $t'$. We use $t'$ to indicate the phase of a particle along this orbit. We have seen in sections~\ref{sec:concept} and~\ref{sec:orbfam} that as a particle moves along the perturbed orbit, it undergoes changes in energy and angular momentum due to exchanges with the perturber. Hence, after some time $\Delta t$, a particle starting from phase $t'$ has transferred a net amount of energy $\Delta E(\Delta t) = E(t' + \Delta t) - E(t')$ to the perturber. Here $E(t')$ is the perturbed energy of a particle at phase $t'$, given by equation~(\ref{Eperturbed}). To work out the total energy exchanged with the perturber by all stars associated with the orbit in question, we need to integrate $\Delta E(\Delta t)$ along the orbit, weighted by the relative number of stars at each point along the orbit. This weight is given by $f_0(E_{0\rmG}(t'))$, with $f_0$ the unperturbed DF, and

\begin{align}
E_{0\rmG}(t')=\frac{1}{2}{\left|\dot{\br}+\bf\Omega_{\rm{\bP}}\times \br-\bv_{\rm\bG}\right|}^2+\Phi_\rmG
\end{align}
the galactocentric energy of the star at phase $t'$ in absence of the perturber, where $\bv_{\rm\bG}=-\Omega_\rmP\, q_\rmG R\, \hat{y}$ is the circular velocity of the galactic center about the COM. If we use $s(t')$ to parameterize the path-length along the phase-space trajectory traced out by the orbit, then the total energy exchanged with the perturber along this orbit, some time $\Delta t$ after the perturber was introduced, is given by the following line-integral
\begin{equation}
\Delta E(\Delta t) = \frac{1}{\calA} \int_s \rmd s(t') \, \left[E(t'+\Delta t)-E(t')\right] \, f_0(E_{0\rmG}(t'))\,.
\label{deltaElineintegral}
\end{equation}
with $\calA$ a normalization factor (see below).

Typically, an orbit in the co-rotating frame will not be exactly closed and the integration limit therefore will have no boundaries. However, for the NCRR orbits discussed in Section~\ref{sec:orbfam}, the orbit is {\it approximately} periodic in the co-rotating frame, with a period $T_{\rm lib}$ set by the time it takes the particle to librate about its Lagrange point (the COC in column~6 of Table~\ref{tab:Ej}), which we compute by a Fourier analysis of the orbit in the co-rotating frame. In the vicinity of the stable Lagrange points, L4 and L5, $T_{\rm lib}$ can be analytically computed using a perturbative method, as discussed in paper II.

The line integral in Eq.~(\ref{deltaElineintegral}) has to be performed along the phase-space trajectory and therefore the differential line element $\rmd s(t')$ is given by $\rmd s = \sqrt{{\left|\rmd \br_i\right|}^2+{\left|\rmd \dot{\br}_i\right|}^2}$. Using that the Jacobian for the transformation from $t'$ to the arc-length $s(t')$ is given by  
\begin{align}
\frac{\rmd s}{\rmd t'} &= \sqrt{{\left|{\dot{\br}}_{\mathrm{\bi}}\right|}^2+{\left|{\ddot{\br}}_{\mathrm{\bi}}\right|}^2}\,,
\end{align}
with ${\dot{\br}}_{\mathrm{\bi}}$ and ${\ddot{\br}}_{\mathrm{\bi}}$ the velocity and acceleration in the inertial frame, respectively, we can approximate the line integral as
\begin{align}
\Delta E(\Delta t) &\approx \frac{1}{\calA} \int_0^{T_{\rm{lib}}} \rmd t' \, \sqrt{{\left|{\dot{\br}}_{\mathrm{\bi}}\right|}^2+{\left|{\ddot{\br}}_{\mathrm{\bi}}\right|}^2}\nonumber \\
&\times \left[E(t'+\Delta t)-E(t')\right] \, f_0(E_{0\rmG}(t'))\,,
\label{deltaE}
\end{align}
with
\begin{align}
\calA &= \int_s \rmd s(t')\,f_0(E_{0\rmG}(t')) \nonumber \\
&= \int_0^{T_{\rm{lib}}} \rmd t' \, \sqrt{{\left|{\dot{\br}}_{\mathrm{\bi}}\right|}^2+{\left|{\ddot{\br}}_{\mathrm{\bi}}\right|}^2} \, f_0(E_{0\rmG}(t'))\,.
\label{norm}
\end{align}
Note that, with this normalization, $\Delta E(\Delta t)$ is the average energy {\it per star} exchanged with the perturber in a time $\Delta t$ along the orbit in question.

The inertial acceleration vector is given by
\begin{align}
{\ddot{\br}}_{\mathrm{\bi}} = -\nabla \Phi\,,
\end{align}
where $\Phi = \Phi_\rmP + \Phi_\rmG$ is the total potential, while the velocity vector in the inertial frame is related to that in the co-rotating frame, 
$\dot{\br}$, by 
\begin{align}
{\dot{\br}}_{\mathrm{\bi}}=\dot{\br}+\bf \Omega_{\rm{\bP}}\times \br\,.
\end{align}

We perform this line integral for the three NCRR orbits (\horseshoen, \pacman and tadpole) shown in Fig.~\ref{fig:orbit2}. All three orbits correspond to our fiducial $q=0.004$ point-mass perturber in a Plummer potential at $R=0.5$. The solid blue, dot-dashed green and dotted red lines in Fig.~\ref{fig:integrated_energy} show the resulting $\Delta E$ for the \horseshoen, \pacman and tadpole orbits respectively as function of $\Delta t$. Note that $\Delta E(\Delta t = T_{\rm lib}) = 0$; as discussed in Section~\ref{sec:concept}, along each NCRR orbit particles both gain and loose energy, and the net effect for a single particle over a full libration period is zero. However, due to the non-uniform phase distribution along each orbit, which arises from the unperturbed phase-space distribution, $f_0(E_{0\rmG})$, we see that $\Delta E$ is positive for all $0 < \Delta t < T_{\rm lib}$. A positive $\Delta E$ indicates that the field particles along these orbits {\it gain} net energy from the perturber, and thus that the perturber experiences dynamical friction. As the field particles gain energy, their $\Omega_\phi$ decreases. The perturber in turn loses energy and falls in, with increasing $\Omega_\rmP$. This puts the original NCRR orbits out of near-co-rotation resonance. Therefore, $\Delta E(\Delta t)$ is only relevant for the dynamics of the system for relatively small $\Delta t$. The exact choice of $\Delta t$ to consider is somewhat ambiguous; it should be indicative of the time scale over which the perturber moves through the resonances, which in turn depends on the strength of dynamical friction. In what follows, we take $\Delta t = T_{\rm orb}$, the orbital time of the perturber. None of our qualitative conclusions are sensitive to this particular choice.

\begin{figure*}[t!]
  \centering
  \begin{subfigure}[t]{0.485\textwidth}
    \centering
    \includegraphics[width=1\textwidth]{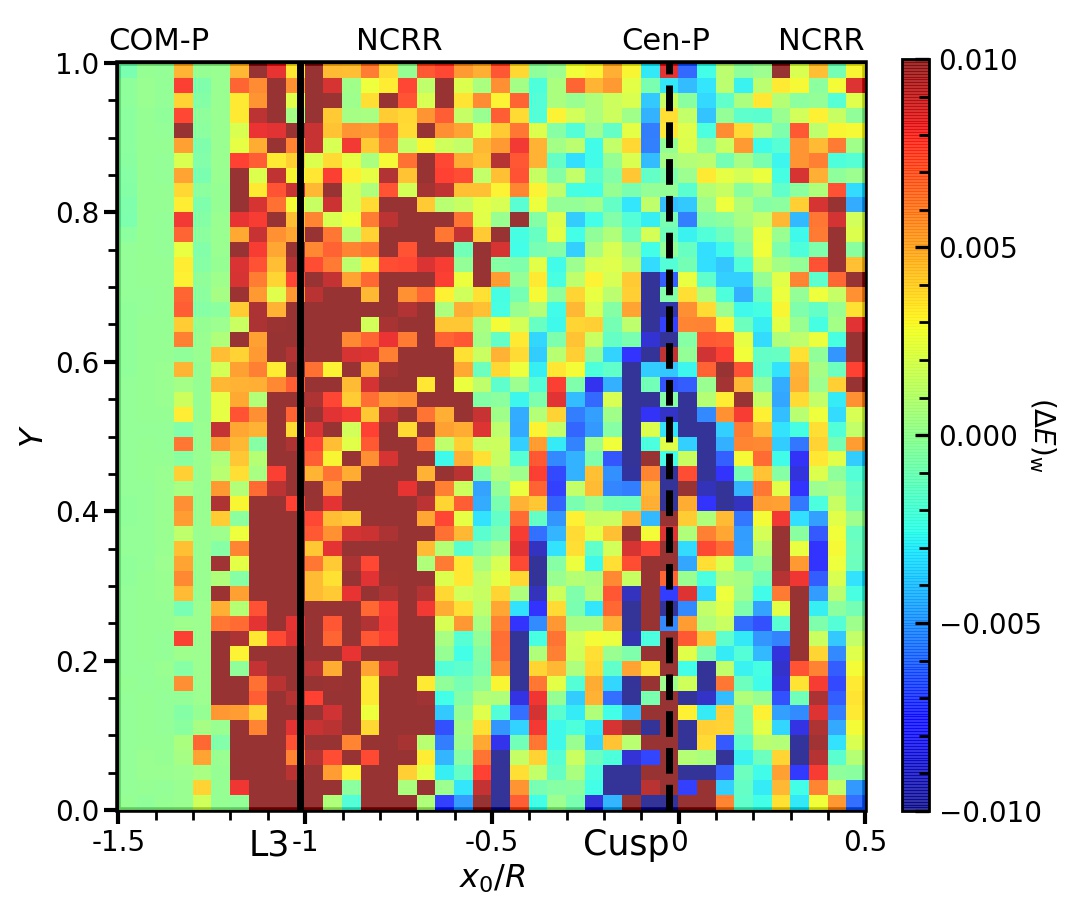}
  \end{subfigure}
  \hspace{2mm}
  \begin{subfigure}[t]{0.463\textwidth}
    \centering
    \includegraphics[width=1\textwidth]{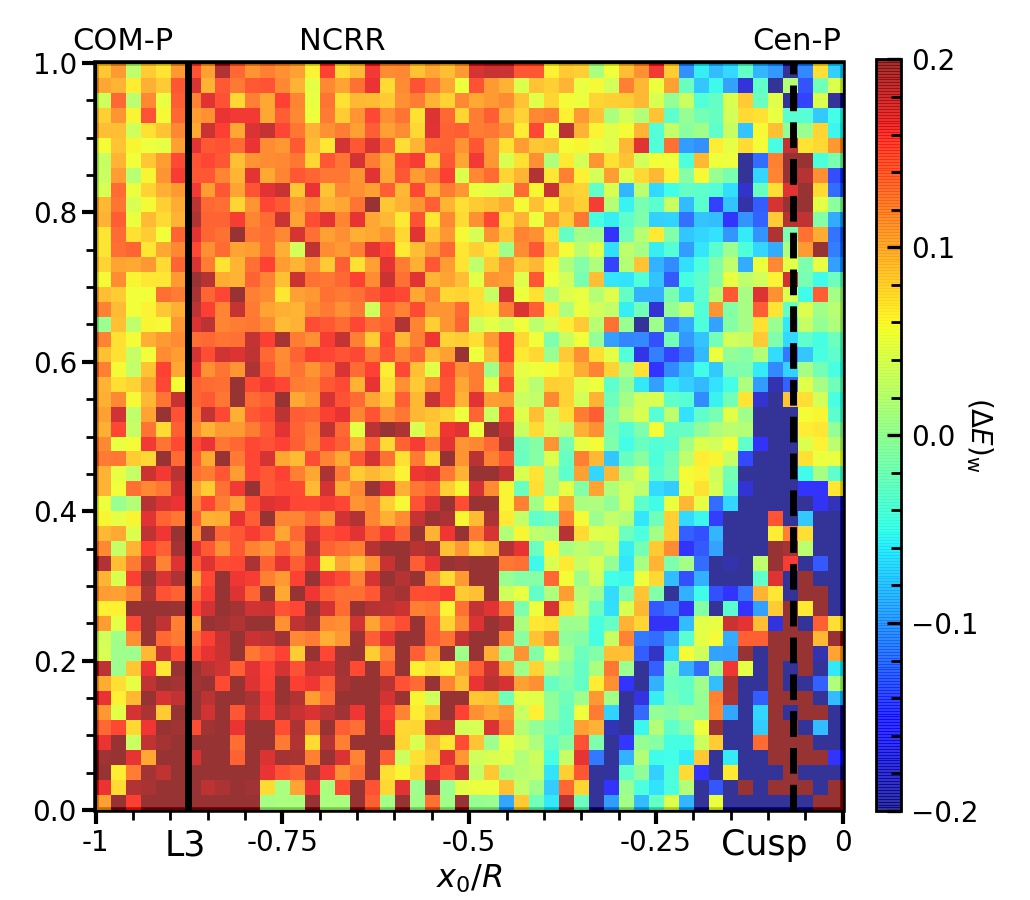}
  \end{subfigure}
  \caption{Same as \ref{fig:delE} but for the cuspy Hernquist potential. Note that $\dE$ is predominantly positive, indicative of a negative (retarding) torque on the perturber. See text for discussion.}
  \label{fig:delE_Hern}
\end{figure*}

The solid blue, dot-dashed green and dotted red curves in Fig.~\ref{fig:integrated_energy} correspond to NCRR orbits in the case where the perturber is orbiting at $R=0.5$, just outside the core of the Plummer sphere. For comparison, the dashed, green curve in Fig.~\ref{fig:integrated_energy} indicates the $\Delta E(\Delta t)$ for a \pacman orbit in the case where the perturber is at $R=0.2$, well inside the core of the Plummer galaxy. In this case $\Delta E$ is negative, indicating that this orbit contributes a positive, enhancing torque. Note that, since the torque on the field particle is given by $\rmd L/\rmd t =  \Omega_\rmP^{-1} (\rmd E/\rmd t)$ (cf. equation~[\ref{dEdt}]), the average torque {\it on the perturber} due to an orbit between $t=0$ and $t=T_{\rm orb}$ is equal to $-\Delta E/\left(\Omega_\rmP T_{\rm orb}\right)=-\Delta E/(2\pi)$, i.e., ${\rm sign}(\calT)=-{\rm sign}(\Delta E)$. We thus see that some of the NCRR orbits can give rise to dynamical buoyancy, rather than friction. An explanation of the latter is discussed in Section~\ref{sec:core}.

\subsection{Scanning Orbital Parameter Space}
\label{sec:scan}

Having demonstrated how to compute the contribution to dynamical friction  from individual orbits, in the form of $\Delta E(T_{\rm orb})$, one can in principle obtain the total torque by summing over all orbits, properly weighted by their relative contribution to the distribution function. In practice, though, this is far from trivial. First of all, sampling all orbits numerically is tedious to the point that one is better off just running an $N$-body simulation. Secondly, some orbits are difficult to integrate accurately, especially some Chimera orbits which reveal semi-ergodic behavior, and the perturber-phylic orbits along which the energy varies rapidly with time. Hence, the non-perturbative method adopted in this paper is not well suited to accurately compute the total dynamical friction torque. Notwithstanding, it gives valuable insight as to the inner workings, in an orbit-based sense, of dynamical friction and buoyancy.

As an example, we now proceed to investigate the contribution to the torque, in terms of $\Delta E(T_{\rm orb})$, from a modest sub-sample of orbits. In what follows we continue to treat the dynamics in 2D (i.e., we only consider orbits in the $x$-$y$ plane depicted in Fig.~\ref{fig:schematic}). We densely sample the part of the orbital parameter space corresponding to the NCRR orbits, which is most relevant for dynamical friction. We first sample the starting point $(x_0,y_0)$ by setting $y_0 = 0$ and sampling $x_0$ uniformly over the range dominated by the NCRR \horseshoe and \pacman orbits (roughly the region inside the $E_\rmJ^{(2)}$ separatrix marked by the solid line in Fig.~\ref{fig:Phi_eff}). Note that by sampling orbits that intersect the $x$-axis, we exclude tadpole orbits with large $E_\rmJ$ that librate in small regions around L4 and L5. After sampling $x_0$, we uniformly sample $E_\rmJ$ over the range $[\Phi_{\rm eff}(x_0,0), E_\rmJ^{(4)}]$. Although orbits with $E_\rmJ\gg E_\rmJ^{(4)}$ are far from co-rotation resonance, thereby contributing less to dynamical friction, those with small, positive values of $E_\rmJ-E^{(4)}_\rmJ$ are NCRR and have similar contribution to the torque as those with $E_\rmJ \lesssim E^{(4)}_\rmJ$. Therefore we consider $E^{(4)}_\rmJ$ to be only an approximate rather than a hard cut-off for the NCRR orbits. Finally, we sample the initial velocities, $v_{\rm x,0}$ and $v_{\rm y,0}$, under the constraint that
\begin{align}
 v_0 = \sqrt{v^2_{\rm x,0} + v^2_{\rm y,0}} = \sqrt{2[E_\rmJ - \Phi_{\rm eff}(x_0,0)]}\,.
\end{align}
Note that both $\bx_0$ and $\bv_0$ are defined in the co-rotating frame. We numerically integrate the orbits for $100\, T_{\rm orb}$, with $T_{\rm orb}$ the orbital time of the perturber, after which we estimate the libration time, $T_{\rm lib}$, by noting the consecutive time-stamps at which each orbit crosses the abscissa of its center-of-circulation (see Table~\ref{tab:Ej}) after making a $2\pi$ circulation about it. Finally, we compute $\Delta E \equiv \Delta E(T_{\rm orb}|x_0, E_\rmJ, v_{\rm x,0}, v_{\rm y,0})$ using equation~(\ref{deltaE}). 

In order to allow for a meaningful comparison of the torque contribution from each of these orbits, we weight the $\Delta E$ {\it per star}, given by equations~(\ref{deltaE})-(\ref{norm}), by the average phase space density associated with that orbit. This yields the total energy exchange per unit phase space from an orbit, given by
\begin{align}
\dE & \equiv \frac{\int_s \rmd s(t') f_0(E_{0\rmG}(t'))}{\int_s \rmd s(t')}\,\Delta E \,.
\end{align}

Using that the time-averaged torque (per unit phase space) on the perturber contributed by an individual orbit is given by
\begin{align}
 \calT_\rmw = -\frac{1}{\Omega_\rmP} \, \frac{\dE}{\Delta t}
\end{align}
(cf. equation~[\ref{dEdt}]), we have that the torque per unit phase space contributed by the orbit can be expressed as
\begin{align}
&\calT_\rmw = -\frac{1}{2\pi}\\
&\times \frac{\int_0^{T_{\rm{lib}}} \rmd t' \, \sqrt{{\left|{\dot{\br}}_{\mathrm{\bi}}\right|}^2+{\left|{\ddot{\br}}_{\mathrm{\bi}}\right|}^2} \left[E(t'+\Delta t)-E(t')\right] f_0(E_{0\rmG}(t'))}{\int_0^{T_{\rm{lib}}} \rmd t' \, \sqrt{{\left|{\dot{\br}}_{\mathrm{\bi}}\right|}^2+{\left|{\ddot{\br}}_{\mathrm{\bi}}\right|}^2}},
\end{align}
where we have used the fact that we adopt $\Delta t = T_{\rm orb} = 2 \pi/\Omega_\rmP$, and we have rewritten $\dE$ using equations~(\ref{deltaE}) and (\ref{norm}).

Fig.~\ref{fig:delE} plots $\dE$ for the Plummer sphere as a function of $x_0$ and $E_\rmJ$ for $\left|v_{\rm y,0}\right| = \frac{1}{3}v_0$. Results for other values of $\left|v_{\rm y,0}\right|$ are very similar, but with the overall amplitudes in $\dE$ decreasing as $\left|v_{\rm y,0}\right| \to v_0$. For each $\left(x_0,E_\rmJ,\left|v_{\rm y,0}\right|\right)$, there are four combinations of $(v_{\rm x,0},v_{\rm y,0})$, given by $(\pm \sqrt{v^2_0 - v^2_{\rm y,0}}, \pm \left|v_{\rm y,0}\right|)$. The values of $\dE$ shown are the sums of these four cases combined.

Left- and right-hand panels correspond to $R=0.5$ and $R=0.2$, respectively. They show the results for a total of $4 \times 2,500$ different orbits. Redder colors denote more positive values of $\dE$ (i.e., stronger dynamical friction), while bluer colors indicate more negative values (i.e., more pronounced dynamical buoyancy). Note that for $R=0.5$, i.e., when the perturber is outside the core, $\dE$ is predominantly positive, indicating that nearly all the NCRR orbits (\horseshoens, \pacmans and some tadpoles, with $x_0$ on either side of L3 and L1) exert a retarding torque (i.e., dynamical friction). However, when $R=0.2$ and the perturber is orbiting inside the core, almost the entire orbital parameter space (dominated by the NCRR \pacmans and tadpoles) contributes to dynamical buoyancy (i.e., $\dE < 0$). Clearly, there is a profound transition in the total torque once the perturber enters the core.

When the perturber is outside the core (left-hand panel), the contribution from the center-phylic orbits, which occupy the range of $x_0$ on either side of the galactic center (L0, marked by the vertical, dashed line) is completely negligible. The same holds for the COM-phylic orbits near the left-most edge of the plot. When the perturber is inside the core (right-hand panel), one again sees that orbits with starting positions close to L0 contribute a negligible torque. Unlike in the left-hand panel, though, these are not center-phylic orbits. After all, those vanish when the perturber crosses the bifurcation radius. Rather, these are predominantly \pacman and tadpole orbits, but unlike their counterparts with starting positions a bit further away from the (unstable) L0, they happen to exert negligible torque. Note that some of the COM-phylic orbits with $x_0/R \lta -0.75$ also contribute a (positive) torque. Their net contribution, though, is significantly smaller than that from the NCRR \pacman orbits, and rapidly weakens when $x_0/R$ becomes smaller (i.e., further away from the galactic center).

Fig.~\ref{fig:delE_Hern} is the same as Fig.~\ref{fig:delE}, but for our Hernquist galaxy. For both $R=0.5$ (left-hand panel) and $R=0.2$ (right-hand panel), it is clear that the total torque is negative (retarding) and dominated by the NCRR orbits. Most importantly, there is no transition in the sign of the total torque as one approaches the center, consistent with the notion that buoyancy and core-stalling are absent if the central density profile is cuspy.
Another difference with respect to the Plummer sphere is that while there is no significant contribution to the torque from the COM-phylic orbits, neither for $R=0.5$, nor for $R=0.2$, the  center-phylic orbits now make a significant contribution to the total torque. Although each of these orbits has a very small $\Delta E(T_{\rm orb})$, the steepness of the distribution function towards the galactic center means that they are abundant, thus receiving a large weight. When the perturber is at $R=0.5$, there are roughly equal numbers of center-phylic orbits with positive and negative $\dE$ (note the alternating red and blue stripes on either side of the galactic center). As a consequence, the net torque contribution from the entire population of center-phylic orbits is small.

Finally, we emphasize that the above inventory of the torque from individual orbits is incomplete. First of all, we have restricted the range of $x_0$ such that it does not include any perturber-phylic orbits. The reason is that they are difficult to integrate, while their contribution to the torque is negligible for reasons discussed in Section~\ref{sec:orbfam}. Secondly, by only picking starting points along the $x$-axis, we have selected against tadpole orbits with large $E_\rmJ$, which are typically confined to small regions centered on L4 or L5. We have examined several of such orbits and found their behavior to be very similar to that of the \horseshoe and \pacman orbits in terms of their contribution to the torque. Thirdly, we have restricted the $E_\rmJ$ values of the orbits up to $E_\rmJ^{(4)}=E_\rmJ^{(5)}$. This is because orbits with $E_\rmJ\gg E_\rmJ^{(4)}$ are far from co-rotation resonance (with drift time steeply falling with increasing $E_\rmJ$) and consequently less important for dynamical friction. However, orbits with small, positive values of $E_\rmJ-E^{(4)}_\rmJ$ have similar contribution to the torque as those with $E_\rmJ\lesssim E^{(4)}_\rmJ$. Thus we use $E^{(4)}_\rmJ$ only as an approximate cut-off for the NCRR orbits. Finally, and most significantly, we have only considered orbits of field particles confined to the orbital plane of the perturber, i.e., those with $z=0$ and $v_\rmz=0$. We presume that this doesn't significantly impact any of our conclusions regarding the contributions of the NCRR \horseshoe and \pacman orbits, as the third dimension merely allows for an additional vertical oscillation not accounted for in our 2D planar treatment (in particular, no new orbital families are introduced by allowing motion in the $z$-direction since there exist no Lagrange points off the orbital plane). However, the relative contributions of the different NCRR orbits to the total torque may be significantly different from what emerges from the 2D analysis presented here. In particular, the tadpole orbits would dominate the phase space and therefore might contribute more significantly to the overall torque in 3D. This is a caveat of our approach that we leave for future work.

\begin{figure*}[t!]
  \centering  \includegraphics[width=1\textwidth]{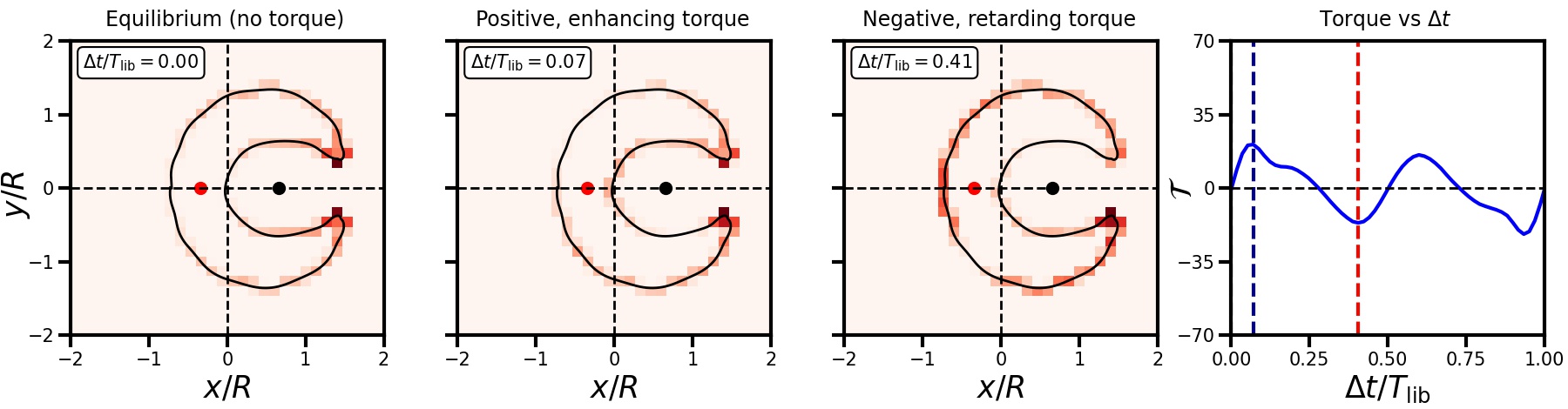}
  \caption{\small Same as Fig.~\ref{fig:wake}, but for a \pacman orbit when the perturber is inside of the core region ($R=0.2$). The first panel from the left shows the unperturbed phase distribution that exerts no torque. In the second panel (corresponding to $\Delta t$ marked by the blue dashed line in the rightmost panel showing the time evolution of the torque) one can note overdensities along the orbit in quadrants I and  III that are responsible for a positive, enhancing torque on the perturber + galactic center. In the third panel (corresponding to $\Delta t$ marked by the red dashed line in the rightmost panel) similar overdensities can be noted in quadrants II and IV, resulting in a negative, retarding torque. Note that the initial torque from this orbit is positive/enhancing, indicating that it will contribute to dynamical buoyancy on the perturber.}
  \label{fig:dynamical_buoyancy}
\end{figure*}

\section{Dynamical buoyancy and core-stalling} 
\label{sec:core}

When the perturber approaches the core region, a bifurcation of some of the Lagrange points causes a drastic change in the orbital structure. As we discuss in detail in paper II, the L3, L0 and L1 points undergo bifurcation at a certain radius $\Rbif$ ($\approx 0.39$ for our fiducial Plummer galaxy plus $q=0.004$ perturber), in which L1 and L3 are annihilated and L0 changes its stability from a center to a saddle. This is associated with the disappearance of the NCRR \horseshoe orbits. The torque from the remaining NCRR \pacman orbits changes from being retarding and contributing to dynamical friction, to being enhancing and contributing to dynamical buoyancy. In this section we discuss why it is that the \pacman orbits (and to some extent also the tadpole orbits) suddenly change the sign of their torque.

When the perturber is well beyond the core radius, the \pacman and tadpole orbits drain energy and angular momentum from the perturber in the same way as the \horseshoe orbits. As described in Section~\ref{sec:concept}, due to the large, radial gradient in the density profile outside of the core, the number density of field particles along the inner section (part of the orbit inside the perturber's radius) of these orbits, which is closer to the galactic center, is larger than that along the outer section (part of the orbit outside the perturber's radius). Due to the clockwise drift motion (in the co-rotating frame), the overdensity along the inner section shifts to the region behind the perturber, creating a `wake' that exerts a retarding torque. This in turn causes the perturber to experience dynamical friction, and thus to move radially inwards (see Fig.~\ref{fig:wake}).

When the perturber is inside the core radius, this picture changes profoundly. The unperturbed galaxy density profile is now very shallow and therefore there no longer is a sharp density contrast of field particles between the inner and outer sections. The equilibrium distribution of particles along the orbit is now dominated by the Jacobian $\sqrt{{\left|{\dot{\br}}_{\mathrm{\bi}}\right|}^2+{\left|{\ddot{\br}}_{\mathrm{\bi}}\right|}^2}$ rather than by the unperturbed distribution function $f_0(E_{0\rmG})$ (cf., equation~[\ref{deltaE}]). And since the particles speed up while approaching the perturber and slow down while receding from it, an overcrowding of particles develops around the inter-section junctions above and below the perturber, as shown in the leftmost panel of Fig.~\ref{fig:dynamical_buoyancy}. As the particles drift along the orbit in a clockwise direction, the overdensity ahead of the perturber approaches it and spreads over the inner section while that behind the perturber moves further away onto the outer section (see the second panel from the left in Fig.~\ref{fig:dynamical_buoyancy}). Hence, contrary to the \horseshoe orbit shown in Fig.~\ref{fig:wake}, here an overdensity of particles first forms {\it ahead} of the perturber, exerting a positive, enhancing torque (marked by the blue dashed line in the rightmost panel that shows the time evolution of the torque) which implies that this orbit will give rise to dynamical buoyancy.

When the perturber is inside the core, some of the tadpole orbits exhibit a similar behavior as the \pacman orbits, thereby contributing to dynamical buoyancy due to orbital-phase-crowding that gives rise to an enhancing torque. The perturber ultimately stalls at a radius where the buoyancy from these orbits is balanced out by friction from the others. As we show in paper II, core-stalling occurs near the bifurcation radius.

\section{Summary}
\label{sec:concl}

Numerical simulations have shown that dynamical friction becomes inefficient inside constant density cores, causing the inward motion of a massive object (the perturber) to stall near the core radius. Objects placed inside this stalling radius are furthermore found to experience dynamical buoyancy that pushes them out towards the stalling radius. This phenomenology is neither predicted by Chandrasekhar's treatment of dynamical friction, nor by the more sophisticated linear, perturbative treatments of TW84 and KS18. The latter infer that dynamical friction arises from the LBK torque due to purely resonant orbits that is exclusively retarding. In BB21, we demonstrated that the LBK torque provides an incomplete description of dynamical friction, which is especially acute in cored galaxies. In particular, we derived an expression for the `self-consistent torque', which includes a memory term that depends on the entire in-fall history of the perturber. As the perturber approaches a core, this memory term causes the net torque to flip sign, i.e., become enhancing, inside a critical radius, $R_{\rm crit}$. Although this formalism thus seems to offer a natural explanation for core stalling, in terms of a balance between friction and buoyancy, it is still based on linear perturbation theory which, as discussed in Section~\ref{sec:intro}, is not justified in the core region of a galaxy, where the perturber can no longer be treated as a weak perturbation, or whenever the perturber does not sweep through the resonances fast enough to prevent non-linearities from building up, i.e., when core-stalling takes effect. 

In this paper, in an attempt to overcome these conceptual problems, we have examined dynamical friction using an alternative, non-perturbative, orbit-based approach. This paints a view of dynamical friction that is subtly different from the standard resonant picture developed in TW84 and KS18. Interactions between the perturber and field particles cause the frequencies (and actions) of the field particles to evolve with time, to the extent that one can no longer talk about particles that obey a commensurability condition (i.e., are in resonance with the perturber) throughout their orbital evolution; rather they are trapped/librating about resonances. As such, dynamical friction does not arise from resonances per se, but rather from an imbalance between the number of particles that are `up-scattered' in angular momentum (or energy) versus those that are `down-scattered' along {\it near-co-rotation-resonant orbits}. This imbalance owes its origin to a non-zero gradient in the distribution function.

We have investigated the inner workings of dynamical friction, on an orbit-by-orbit basis, for the case of a point mass perturber on a circular orbit at radius $R$ in a spherical host galaxy. By assuming that $|\rmd R/\rmd t|$ of the perturber is small, i.e., we are in the `slow' regime, appropriate for studying core stalling, the motion of the field particles can be treated as a restricted three-body problem in which the Jacobi energy of the field particles is conserved.  Each individual orbit in the fully perturbed, time-dependent potential of the galaxy$+$perturber system has phases at which the orbital angular momentum and energy increase (the perturber experiences friction) and decrease (the perturber experiences buoyancy). Since the Jacobi energy is conserved, the net effect of these energy changes, when integrated over a full libration period in the frame co-rotating with the perturber, is nullified. For dynamical friction to emerge, then, two conditions need to be satisfied: (i) there need to be orbits with a non-uniform phase distribution along the orbit for which the time-lag between the two phases corresponding to retarding (friction) and enhancing (buoyancy) torques is sufficiently long, and (ii) there needs to be sufficient phase-coherence among different orbits. In that case, if all these phase-coherent orbits first exert friction on the perturber, the latter can sink in, modifying its frequency significantly, before the orbits would enter their buoyancy-exerting phase.

We have shown that dynamical friction is dominated by orbits that have (unperturbed) azimuthal frequencies similar to the circular frequency of the perturber. These near-co-rotation-resonant (NCRR) orbits all have a long libration time in the frame co-rotating with the perturber, assuring a long time-lag between the orbit's contribution to a retarding torque (friction) and an enhancing torque (buoyancy). And since all NCRR orbits have the same sense of rotation in the co-rotating frame, phase-coherence is guaranteed. Other orbits, such as the center- or perturber-phylic ones have a time-lag between the friction and buoyancy phases that is too short for a net, coherent torque to emerge. In other cases, especially the COM-phylic orbits, the phase-density along the orbit is almost uniform, such that again no net torque arises.

We have identified three different families of NCRR orbits that dominate the contribution to dynamical friction: \horseshoe orbits that circulate the Lagrange point L3, tadpole orbits that librate around either L4 or L5, and \pacman orbits which circulate the galactic center and pass through the region between the equipotential contours corresponding to L1 and L2. \Horseshoe and tadpole orbits are relatively well-known in planetary dynamics \citep[e.g.,][]{Dermott.Murray.81, Goldreich.Tremaine.82}. For example, objects on tadpole orbits are known as trojans, which includes large swarms of trojan asteroids associated with Jupiter, as well as several trojan moons in the Saturn system. There are several asteroids known to be on \horseshoe orbits in the Earth-Sun system \citep[][]{Connors.etal.02, Brasser.etal.04, Christou.Asher.11}, while the Saturnian moons Janus and Epimetheus are known to be horse-shoeing each other \citep[][]{Dermott.Murray.81b}. The \horseshoe and tadpole orbits are also key players in galactic dynamics, where they are often referred to as `trapped' orbits. They play a key role in phenomena such as the radial migration of stars in disk galaxies induced by perturbations due to a bar or spiral arm \citep[e.g.,][]{Barbanis.76, Carlberg.Sellwood.85, Sellwood.Binney.02, Daniel.Wyse.15}. However, to our knowledge, the orbits that we have called \pacman orbits, because of their characteristic shape, have hitherto not been identified as a separate orbital class. Pacman orbits are only present if the galaxy has a cored density profile ($\rmd\log\rho/\rmd\log r>-1$), in which case, the center of the galaxy is a stationary Lagrange point, which we have dubbed L0.

In a cusp, \pacman orbits are absent, and dynamical friction is mainly caused by field particles moving along the NCRR \horseshoe and tadpole orbits. Due to a large, negative gradient in the distribution function, the vast majority of these orbits yield a net retarding torque, draining orbital energy and angular momentum from the perturber.

In a cored profile, the behavior is very different. Well outside the core, where the density gradient is steep, the NCRR \horseshoen, tadpole and \pacman orbits exert a retarding torque, just as in the case of a cuspy density profile.  However, as the perturber enters the core region, the orbital configuration changes drastically. First of all, as we show in Paper~II, the  Lagrange points L3, L1 and L0 undergo a bifurcation in which L3 and L1 are annihilated, while L0 changes from a stable center to an unstable saddle. As a consequence, the \horseshoe orbits disappear, making the tadpoles and \pacmans the only surviving NCRR orbits. This disappearance of the \horseshoe orbits is equivalent to the suppression of low-order resonances in the core region, advocated by KS18 as the main cause of core stalling. However, we have demonstrated that a large number of NCRR orbits (\pacmans and tadpoles) remain, which continue to exchange energy and angular momentum with the perturber. The pre-eminent cause of core stalling, therefore, is not the disappearance of resonances, but the fact that most of the remaining \pacman and tadpole orbits now give rise to a net enhancing torque, thereby effectuating `dynamical buoyancy'. The main reason for this reversal in the sign of the net torque is the dramatic change in the radial gradient of the density distribution as described in Section~\ref{sec:core}. 

With dynamical buoyancy dominating over dynamical friction in the central region of a cored density profile, the perturber will ultimately settle at a core-stalling radius where the outward buoyant force balances friction. This notion of buoyancy counteracting friction in the core region is supported by numerical simulations \citep[][]{Inoue.11, Cole.etal.12, Petts.etal.16}, and provides a natural explanation for core stalling. In Paper~II we show that core-stalling happens close to a critical `bifurcation radius', $R_{\rm bif}$.

Dynamical buoyancy has a number of important astrophysical implications. It can prevent massive objects like black holes, globular clusters, and satellite galaxies from sinking all the way to the center of their host system, if the latter has a central constant density core. Hence, buoyancy acts as a natural barrier for, among others, the merging of supermassive black holes (SMBHs), with implications for the expected rates of such events to be detected by future gravitational wave detectors such as LISA \citep[e.g.,][]{Rhook.Wyithe.05, Tremmel.etal.18, Ricarte.Natarajan.18}, and for the creation of nuclear star clusters through the merging of globular clusters \citep[e.g.,][]{Tremaine.etal.75, ArcaSedda.CapuzzoDolcetta.17, Boldrini.etal.19}. Put differently, if their formation mechanism is merger driven, then the presence of central SMBHs and/or nuclear star clusters would favor cuspy density profiles for their hosts, which could help to constrain the particle nature of dark matter \citep[e.g.,][and references therein]{Brooks.14}.

However, many outstanding issues remain. For example, the analysis presented here has largely focused on orbits in 2D, and needs to be extended to 3D. It is also important to examine how friction and buoyancy act on perturbers along non-circular orbits and/or in non-spherical potentials, both of which are expected to result in a much richer dynamics \citep[e.g.,][]{Capuzzo-Dolcetta.Vicari.05}. In our analysis we also neglected the radial motion of the perturber due to friction/buoyancy itself. Although a reasonable approximation to make for studying dynamical friction in the `slow' regime, especially core-stalling, BB21 have shown that the memory effect of dynamical friction, i.e., the dependence on the perturber's past in-fall-history, can play an important role, something that warrants further investigation within the non-perturbative framework presented here. And finally, more work is needed to assess if and how core-stalling depends on the central, logarithmic slope, $\gamma$, of the host galaxy.  In this paper we have focused exclusively on two special cases; a constant density core with $\gamma = 0$, and a steep NFW-like cusp with $\gamma = -1$. Numerical simulations suggest that core stalling might be present as long as $\gamma > -1$ \citep[][]{Goerdt.etal.10}. In future work we intend to examine dynamical friction and core stalling in host-galaxies with a variety of different central density slopes in the range $-1 \leq \gamma \leq 0$, using a combination of numerical simulations and the orbit-based formalism presented here. 

\section*{Acknowledgments}

The authors are grateful to the anonymous referee for an extremely detailed and insightful report, and to Martin Weinberg, Seshadri Sridhar, Dhruba Dutta-Chowdhury, Nir Mandelker and Kaustav Mitra for valuable discussions. FvdB is supported by the National Aeronautics and Space Administration through Grant No. 19-ATP19-0059 issued as part of the Astrophysics Theory Program, and received additional support from the Klaus Tschira foundation. The work of FvdB was performed in part at the Aspen Center for Physics, which is supported by the National Science Foundation grant PHY-1607611.


\bibliography{references_vdb}{}

\begin{thebibliography}{}
\expandafter\ifx\csname natexlab\endcsname\relax\def\natexlab#1{#1}\fi
\providecommand{\url}[1]{\href{#1}{#1}}
\providecommand{\dodoi}[1]{doi:~\href{http://doi.org/#1}{\nolinkurl{#1}}}
\providecommand{\doeprint}[1]{\href{http://ascl.net/#1}{\nolinkurl{http://ascl.net/#1}}}
\providecommand{\doarXiv}[1]{\href{https://arxiv.org/abs/#1}{\nolinkurl{https://arxiv.org/abs/#1}}}

\bibitem[{{Arca-Sedda} \&
  {Capuzzo-Dolcetta}(2017)}]{ArcaSedda.CapuzzoDolcetta.17}
{Arca-Sedda}, M., \& {Capuzzo-Dolcetta}, R. 2017, \mnras, 464, 3060,
  \dodoi{10.1093/mnras/stw2483}

\bibitem[{{Athanassoula} {et~al.}(1983){Athanassoula}, {Bienayme}, {Martinet},
  \& {Pfenniger}}]{Athanassoula.etal.83}
{Athanassoula}, E., {Bienayme}, O., {Martinet}, L., \& {Pfenniger}, D. 1983,
  \aap, 127, 349

\bibitem[{{Banik} \& {van den Bosch}(2021)}]{Banik.vdBosch.2021}
{Banik}, U., \& {van den Bosch}, F.~C. 2021, \apj, 912, 43,
  \dodoi{10.3847/1538-4357/abeb6d}

\bibitem[{{Barbanis}(1976)}]{Barbanis.76}
{Barbanis}, B. 1976, Celestial Mechanics, 14, 201, \dodoi{10.1007/BF01376320}

\bibitem[{{Binney} \& {Tremaine}(1987)}]{Binney.Tremaine.87}
{Binney}, J., \& {Tremaine}, S. 1987, {Galactic dynamics} (Princeton University
  Press)

\bibitem[{{Binney} \& {Tremaine}(2008)}]{Binney.Tremaine.08}
---. 2008, {Galactic Dynamics: Second Edition} (Princeton University Press)

\bibitem[{{Boldrini} {et~al.}(2019){Boldrini}, {Mohayaee}, \&
  {Silk}}]{Boldrini.etal.19}
{Boldrini}, P., {Mohayaee}, R., \& {Silk}, J. 2019, \mnras, 485, 2546,
  \dodoi{10.1093/mnras/stz573}

\bibitem[{{Boylan-Kolchin} {et~al.}(2008){Boylan-Kolchin}, {Ma}, \&
  {Quataert}}]{Boylan-Kolchin.etal.08}
{Boylan-Kolchin}, M., {Ma}, C.-P., \& {Quataert}, E. 2008, \mnras, 383, 93,
  \dodoi{10.1111/j.1365-2966.2007.12530.x}

\bibitem[{{Brasser} {et~al.}(2004){Brasser}, {Innanen}, {Connors}, {Veillet},
  {Wiegert}, {Mikkola}, \& {Chodas}}]{Brasser.etal.04}
{Brasser}, R., {Innanen}, K.~A., {Connors}, M., {et~al.} 2004, \icarus, 171,
  102, \dodoi{10.1016/j.icarus.2004.04.019}

\bibitem[{{Brooks}(2014)}]{Brooks.14}
{Brooks}, A. 2014, Annalen der Physik, 264, 294, \dodoi{10.1002/andp.201400068}

\bibitem[{{Capuzzo-Dolcetta} \& {Vicari}(2005)}]{Capuzzo-Dolcetta.Vicari.05}
{Capuzzo-Dolcetta}, R., \& {Vicari}, A. 2005, \mnras, 356, 899,
  \dodoi{10.1111/j.1365-2966.2004.08433.x}

\bibitem[{{Carlberg} \& {Sellwood}(1985)}]{Carlberg.Sellwood.85}
{Carlberg}, R.~G., \& {Sellwood}, J.~A. 1985, \apj, 292, 79,
  \dodoi{10.1086/163134}

\bibitem[{{Chandrasekhar}(1943)}]{Chandrasekhar.43}
{Chandrasekhar}, S. 1943, \apj, 97, 255, \dodoi{10.1086/144517}

\bibitem[{{Chiba} \& {Sch{\"o}nrich}(2021)}]{Chiba.Schonrich.21}
{Chiba}, R., \& {Sch{\"o}nrich}, R. 2021, arXiv e-prints, arXiv:2109.10910.
\newblock \doarXiv{2109.10910}

\bibitem[{{Christou} \& {Asher}(2011)}]{Christou.Asher.11}
{Christou}, A.~A., \& {Asher}, D.~J. 2011, \mnras, 414, 2965,
  \dodoi{10.1111/j.1365-2966.2011.18595.x}

\bibitem[{{Cole} {et~al.}(2012){Cole}, {Dehnen}, {Read}, \&
  {Wilkinson}}]{Cole.etal.12}
{Cole}, D.~R., {Dehnen}, W., {Read}, J.~I., \& {Wilkinson}, M.~I. 2012, \mnras,
  426, 601, \dodoi{10.1111/j.1365-2966.2012.21885.x}

\bibitem[{{Connors} {et~al.}(2002){Connors}, {Chodas}, {Mikkola}, {Wiegert},
  {Veillet}, \& {Innanen}}]{Connors.etal.02}
{Connors}, M., {Chodas}, P., {Mikkola}, S., {et~al.} 2002, Meteoritics and
  Planetary Science, 37, 1435, \dodoi{10.1111/j.1945-5100.2002.tb01039.x}

\bibitem[{{Contopoulos}(1973)}]{Contopoulos.73}
{Contopoulos}, G. 1973, \apj, 181, 657, \dodoi{10.1086/152080}

\bibitem[{{Contopoulos}(1975)}]{Contopoulos.75}
---. 1975, \apj, 201, 566, \dodoi{10.1086/153922}

\bibitem[{{Contopoulos}(1979)}]{Contopoulos.79}
---. 1979, {Integrable and stochastic behaviour in dynamical astronomy}, ed.
  G.~{Casati} \& J.~{Ford}, Vol.~93, 1--17, \dodoi{10.1007/BFb0021733}

\bibitem[{{Cora} {et~al.}(1997){Cora}, {Muzzio}, \& {Vergne}}]{Cora.etal.97}
{Cora}, S.~A., {Muzzio}, J.~C., \& {Vergne}, M.~M. 1997, \mnras, 289, 253,
  \dodoi{10.1093/mnras/289.2.253}

\bibitem[{{Daniel} \& {Wyse}(2015)}]{Daniel.Wyse.15}
{Daniel}, K.~J., \& {Wyse}, R. F.~G. 2015, \mnras, 447, 3576,
  \dodoi{10.1093/mnras/stu2683}

\bibitem[{{Dermott} \& {Murray}(1981{\natexlab{a}})}]{Dermott.Murray.81}
{Dermott}, S.~F., \& {Murray}, C.~D. 1981{\natexlab{a}}, \icarus, 48, 1,
  \dodoi{10.1016/0019-1035(81)90147-0}

\bibitem[{{Dermott} \& {Murray}(1981{\natexlab{b}})}]{Dermott.Murray.81b}
---. 1981{\natexlab{b}}, \icarus, 48, 12, \dodoi{10.1016/0019-1035(81)90148-2}

\bibitem[{{Dutta Chowdhury} {et~al.}(2019){Dutta Chowdhury}, {van den Bosch},
  \& {van Dokkum}}]{DuttaChowdhury.etal.19}
{Dutta Chowdhury}, D., {van den Bosch}, F.~C., \& {van Dokkum}, P. 2019, \apj,
  877, 133, \dodoi{10.3847/1538-4357/ab1be4}

\bibitem[{{Fouvry} \& {Bar-Or}(2018)}]{Fouvry.Bar-Or.18}
{Fouvry}, J.-B., \& {Bar-Or}, B. 2018, \mnras, 481, 4566,
  \dodoi{10.1093/mnras/sty2571}

\bibitem[{{Goerdt} {et~al.}(2010){Goerdt}, {Moore}, {Read}, \&
  {Stadel}}]{Goerdt.etal.10}
{Goerdt}, T., {Moore}, B., {Read}, J.~I., \& {Stadel}, J. 2010, \apj, 725,
  1707, \dodoi{10.1088/0004-637X/725/2/1707}

\bibitem[{{Goldreich} \& {Tremaine}(1982)}]{Goldreich.Tremaine.82}
{Goldreich}, P., \& {Tremaine}, S. 1982, \araa, 20, 249,
  \dodoi{10.1146/annurev.aa.20.090182.001341}

\bibitem[{{Hashimoto} {et~al.}(2003){Hashimoto}, {Funato}, \&
  {Makino}}]{Hashimoto.etal.03}
{Hashimoto}, Y., {Funato}, Y., \& {Makino}, J. 2003, \apj, 582, 196,
  \dodoi{10.1086/344260}

\bibitem[{{Henon}(1959)}]{Henon.59}
{Henon}, M. 1959, Annales d'Astrophysique, 22, 126

\bibitem[{{Henrard}(1982)}]{Henrard.82}
{Henrard}, J. 1982, Celestial Mechanics, 27, 3, \dodoi{10.1007/BF01228946}

\bibitem[{{Hernandez} \& {Gilmore}(1998)}]{Hernandez.Gilmore.98}
{Hernandez}, X., \& {Gilmore}, G. 1998, \mnras, 297, 517,
  \dodoi{10.1046/j.1365-8711.1998.01511.x}

\bibitem[{{Hernquist}(1990)}]{Hernquist.90}
{Hernquist}, L. 1990, \apj, 356, 359, \dodoi{10.1086/168845}

\bibitem[{{Inoue}(2011)}]{Inoue.11}
{Inoue}, S. 2011, \mnras, 416, 1181, \dodoi{10.1111/j.1365-2966.2011.19122.x}

\bibitem[{{Jiang} {et~al.}(2008){Jiang}, {Jing}, {Faltenbacher}, {Lin}, \&
  {Li}}]{Jiang.etal.08}
{Jiang}, C.~Y., {Jing}, Y.~P., {Faltenbacher}, A., {Lin}, W.~P., \& {Li}, C.
  2008, \apj, 675, 1095, \dodoi{10.1086/526412}

\bibitem[{{Kaur} \& {Sridhar}(2018)}]{Kaur.Sridhar.18}
{Kaur}, K., \& {Sridhar}, S. 2018, \apj, 868, 134,
  \dodoi{10.3847/1538-4357/aaeacf}

\bibitem[{Kotovych \& Bowman(2002)}]{Kotovych_2002}
Kotovych, O., \& Bowman, J.~C. 2002, Journal of Physics A: Mathematical and
  General, 35, 7849, \dodoi{10.1088/0305-4470/35/37/301}

\bibitem[{{Lichtenberg} \& {Lieberman}(1992)}]{Lichtenberg.Lieberman.92}
{Lichtenberg}, A., \& {Lieberman}, M. 1992, {Regular and Chaotic Dynamics}

\bibitem[{{Lin} \& {Tremaine}(1983)}]{Lin.Tremaine.83}
{Lin}, D.~N.~C., \& {Tremaine}, S. 1983, \apj, 264, 364, \dodoi{10.1086/160604}

\bibitem[{{Lynden-Bell} \& {Kalnajs}(1972)}]{LyndenBell.Kalnajs.72}
{Lynden-Bell}, D., \& {Kalnajs}, A.~J. 1972, \mnras, 157, 1,
  \dodoi{10.1093/mnras/157.1.1}

\bibitem[{{Petts} {et~al.}(2015){Petts}, {Gualandris}, \&
  {Read}}]{Petts.etal.15}
{Petts}, J.~A., {Gualandris}, A., \& {Read}, J.~I. 2015, \mnras, 454, 3778,
  \dodoi{10.1093/mnras/stv2235}

\bibitem[{{Petts} {et~al.}(2016){Petts}, {Read}, \&
  {Gualandris}}]{Petts.etal.16}
{Petts}, J.~A., {Read}, J.~I., \& {Gualandris}, A. 2016, \mnras, 463, 858,
  \dodoi{10.1093/mnras/stw2011}

\bibitem[{{Plummer}(1911)}]{Plummer.11}
{Plummer}, H.~C. 1911, \mnras, 71, 460, \dodoi{10.1093/mnras/71.5.460}

\bibitem[{{Read} {et~al.}(2006){Read}, {Goerdt}, {Moore}, {Pontzen}, {Stadel},
  \& {Lake}}]{Read.etal.06c}
{Read}, J.~I., {Goerdt}, T., {Moore}, B., {et~al.} 2006, \mnras, 373, 1451,
  \dodoi{10.1111/j.1365-2966.2006.11022.x}

\bibitem[{{Rhook} \& {Wyithe}(2005)}]{Rhook.Wyithe.05}
{Rhook}, K.~J., \& {Wyithe}, J. S.~B. 2005, \mnras, 361, 1145,
  \dodoi{10.1111/j.1365-2966.2005.08987.x}

\bibitem[{{Ricarte} \& {Natarajan}(2018)}]{Ricarte.Natarajan.18}
{Ricarte}, A., \& {Natarajan}, P. 2018, \mnras, 474, 1995,
  \dodoi{10.1093/mnras/stx2851}

\bibitem[{{Sellwood} \& {Binney}(2002)}]{Sellwood.Binney.02}
{Sellwood}, J.~A., \& {Binney}, J.~J. 2002, \mnras, 336, 785,
  \dodoi{10.1046/j.1365-8711.2002.05806.x}

\bibitem[{{Tremaine} \& {Weinberg}(1984)}]{Tremaine.Weinberg.84}
{Tremaine}, S., \& {Weinberg}, M.~D. 1984, \mnras, 209, 729,
  \dodoi{10.1093/mnras/209.4.729}

\bibitem[{{Tremaine} {et~al.}(1975){Tremaine}, {Ostriker}, \&
  {Spitzer}}]{Tremaine.etal.75}
{Tremaine}, S.~D., {Ostriker}, J.~P., \& {Spitzer}, L., J. 1975, \apj, 196,
  407, \dodoi{10.1086/153422}

\bibitem[{{Tremmel} {et~al.}(2018){Tremmel}, {Governato}, {Volonteri}, {Quinn},
  \& {Pontzen}}]{Tremmel.etal.18}
{Tremmel}, M., {Governato}, F., {Volonteri}, M., {Quinn}, T.~R., \& {Pontzen},
  A. 2018, \mnras, 475, 4967, \dodoi{10.1093/mnras/sty139}

\bibitem[{{van den Bosch} {et~al.}(1999){van den Bosch}, {Lewis}, {Lake}, \&
  {Stadel}}]{vdBosch.etal.99}
{van den Bosch}, F.~C., {Lewis}, G.~F., {Lake}, G., \& {Stadel}, J. 1999, \apj,
  515, 50, \dodoi{10.1086/307023}

\bibitem[{{Zelnikov} \& {Kuskov}(2016)}]{Zelnikov.Kuskov.16}
{Zelnikov}, M.~I., \& {Kuskov}, D.~S. 2016, \mnras, 455, 3597,
  \dodoi{10.1093/mnras/stv2389}

\end{thebibliography}
\bibliographystyle{aasjournal}


\appendix

\section{Orbit classification}
\label{App:orb_class}

As shown in \cite{Daniel.Wyse.15}, the Jacobi energy of an NCRR orbit can be obtained by a perturbative expansion around the co-rotation L4/L5 points in terms of angular momentum and radial action, using the third order epicyclic theory of \cite{Contopoulos.75}. We extend this analysis to include orbits in the vicinity of not only L4 and L5, but also L0 and the perturber. About any such stable fixed point, which we shall refer to as the center of perturbation (COP) hereon, the Jacobi energy (defined wrt the galactic center) can be perturbatively expanded up to second order in actions as the following series
\citep[equation~A29 of][]{Contopoulos.75}:
\begin{align}
E'_\rmJ = h'_0 + \Delta\Omega_0 J_\varphi + \kappa_0 J_r + a_0 J^2_r + 2 b_0 J_r J_\varphi + c_0 J^2_\varphi + \Phi_1(r_\rmG,\varphi),
\label{Ej_perturb}
\end{align}
where $r_\rmG$ is the distance from the galactic center and $\varphi$ is the anticlockwise angle measured from the x-axis wrt the galactic center. $h'_0$ is the unperturbed Jacobi energy (wrt the galactic center) evaluated at the COP. The Jacobi energy, $E_\rmJ$, wrt the COM of the galaxy-perturber system is related to that wrt the galactic center, $E'_\rmJ$, as follows

\begin{align}
E_\rmJ = E'_\rmJ + \frac{1}{2}\Omega^2_\rmP q^2_\rmG R^2.
\end{align}
$J_r$ is the radial action (wrt the galactic center), $J_\varphi=L-L_0$ is the angular momentum relative to the COP (with $L=r^2_\rmG\dot{\varphi}$ and $L_0=r^2_{\rmG 0}\Omega_0$, where $r_{\rmG0}$ is the distance of the COP from the galactic center), and $\Delta \Omega_0=\Omega_0-\Omega_\rmP$, with $\Omega_0$ the azimuthal frequency evaluated at the COP. The radial epicyclic frequency, $\kappa_0$, and the constants $a_0$, $b_0$ and $c_0$ are evaluated at the COP in terms of the galaxy potential as follows
\begin{align}
\kappa^2_0 &= \Phi^{''}_\rmG+3\Omega^2_0,\nonumber \\
a_0 &= \frac{1}{16}\kappa^2_0\left[\Phi^{''''}_\rmG+\frac{60\Phi^{'}_\rmG}{r^3_0}-\frac{5}{3\kappa^2_0}{\left(\Phi^{'''}_\rmG-\frac{12\Phi^{'}_\rmG}{r^2_0}\right)}^2\right],\nonumber \\
b_0 &= \frac{\Omega_0\kappa^{'}_0}{r_0\kappa^2_0},\nonumber \\
c_0 &= \frac{\Omega_0\Omega^{'}_0}{r_0\kappa^2_0},
\end{align}
where each prime denotes a derivative with respect to $r_\rmG$. $\Phi_1$ is the disturbing potential that includes the perturber potential $\Phi_\rmP$ and the tidal potential due to the orbital motion of the galactic center about the COM, i.e.,
\begin{align}
\Phi_1(r_\rmG,\varphi) &= \Phi_\rmP + \frac{q\,r_\rmG\cos{\varphi}}{R^2} \nonumber\\
&= q\left[-\frac{1}{\sqrt{r^2_\rmG+R^2-2r_\rmG R\cos{\varphi}}}+\frac{r_\rmG\cos{\varphi}}{R^2}\right].
\label{Phip_tidal}
\end{align}

Following \citet{Daniel.Wyse.15}, equation~(\ref{Ej_perturb}) can be written in the following
quadratic form:
\begin{align}
A J^2_\varphi+B J_\varphi + C = 0,
\end{align}
where $A$, $B$ and $C$ are given by
\begin{align}
A &= c_0,\nonumber \\
B &= \Delta \Omega_0 + 2 b_0 J_r,\nonumber \\
C &= a_0 J^2_r + \kappa_0 J_r + h'_0 + \Phi_1(r,\varphi) - E'_\rmJ.
\end{align}
The particle can only venture into those regions of $(r_\rmG,\varphi)$ where $J_\varphi$ has real solutions, which occurs when
\begin{align}
B^2-4 A C \geq 0,
\end{align}
i.e.,
\begin{align}
E'_{\rm Jc}=E'_\rmJ - \frac{{\left(\Delta\Omega_0\right)}^2}{4\left|c_0\right|} - \left(\kappa_0+\frac{b_0}{\left|c_0\right|}\Delta \Omega_0\right)J_r - \left(a_0+\frac{b^2_0}{\left|c_0\right|}\right)J^2_r \leq h'_0+\Phi_1(r,\varphi).
\label{Ejcirc_app}
\end{align}
Here $E'_{\rm Jc}$ is the circular part of the Jacobi energy (we have used the fact that $c_0=-\left|c_0\right|<0$ for realistic galaxy profiles). The region accessible to the particle, i.e., the range of $\varphi$ for which $J_\varphi$ has real roots, depends on the value of $E'_{\rm Jc}$ relative to that of $\Phi_1$ at the separatrix, the zero-velocity curve (ZVC) passing through the saddle point $(r_{\rm sep},\varphi_{\rm sep})$ nearest to the COP. If the accessible region is `inside' (`outside') the separatrix, i.e., $E'_{\rm Jc}>h'_0+\Phi_1(r_{\rm sep},\varphi_{\rm sep})$ ($E'_{\rm Jc}<h'_0+\Phi_1(r_{\rm sep},\varphi_{\rm sep})$), then the orbit is said to be `trapped' (`untrapped'). However, since $J_r$ can oscillate along an orbit, especially for orbits that are  further away from co-rotation resonance (see \S\ref{sec:slowfast}), the trapping criterion is not guaranteed to be satisfied forever; the field particle can oscillate in and out of the trapped region by crossing the separatrix and transitioning between different orbit-families (see \S\ref{sec:crossing}).

\subsection{Perturbation about L4 and L5}

L4 and L5 are located at a distance $r_\rmG=R$ from the galactic center, and at an angle $\varphi=\pm\pi/3$. The nearest saddle point to L4/L5 is L3, at a distance of $r_\rmG=r_{\rm sep}=r_3$ from the galactic center, and with $\varphi=\varphi_{\rm sep}=\pi$. In the region centred on L4/L5 and bounded by the L3 separatrix, the disturbing potential $\Phi_1$, given by equation~(\ref{Phip_tidal}), is bounded by
\begin{align}
-q\left[\frac{1}{r_3 + R} + \frac{r_3}{R^2}\right] = \Phi_1(r_3,\pi) < \Phi_1 < \Phi_1(R,\pm\pi/3) = -\frac{q}{2 R}.
\end{align}
Therefore, in the vicinity of L4, L5 and L3, $J_\phi$ has real roots for $\varphi$ that are restricted to $0<\varphi<\pi$ (or $-\pi<\varphi<0$) when $h_0^{'(4)}+\Phi_1(r_3,\pi) < E_{\rm Jc}^{'(4)} < h_0^{'(4)} + \Phi_1(R,\pm\pi/3)$, i.e., when the `trapping criterion',
\begin{align}
h^{'(4)}_0-q\left[\frac{1}{r_3 + R}+\frac{r_3}{R^2}\right]=E^{'(3)}_\rmJ < E_{\rm Jc}^{'(4)} < E^{'(4)}_\rmJ=h^{'(4)}_0 - \frac{q}{2 R},
\end{align}
is satisfied. Here $E_{\rm Jc}^{'(4)}$ and $h_0^{'(4)}$ are, respectively, the circular part of the Jacobi energy given by equation~(\ref{Ejcirc_app}) and the unperturbed Jacobi energy, both evaluated at the COP, L4/L5. These `trapped' orbits that lie inside the L3 separatrix are the tadpole orbits. The `untrapped' orbits that lie beyond the L3 separatrix ($\varphi>\varphi_{\rm sep}=\pi$), i.e., that satisfy the condition,
\begin{align}
E_{\rm Jc}^{'(4)} < E^{'(3)}_\rmJ,
\end{align}
are the horse-shoe orbits. The tadpole orbits are trapped, librating around L4/L5, while the \horseshoe orbits are untrapped wrt L4/L5. However, as we shall see shortly, the \horseshoe orbits are trapped inside the L1/L2 separatrix and are therefore still librating about near-co-rotation resonances. Since $J_r$ can vary along an orbit, certain orbits with high $J_r$ can cross the L3 separatrix, and the trapped tadpoles and untrapped \horseshoes can metamorphose into each other, showing Chimera behavior (see Appendix~\ref{App:Chimera}).

\subsection{Perturbation about L0}

A perturbative analysis around L0 can only be performed for a perfect central core, i.e., $\gamma \equiv \lim_{r \to 0}\rmd\log\rho/\rmd\log r=0$. For density profiles with $\gamma<0$, $\Omega_0$ and $\kappa_0$ diverge like $r^{\gamma/2}_\rmG$ towards the galactic center, around which the perturbative expansion of $E'_\rmJ$ given in equation~(\ref{Ej_perturb}) is thus not defined. 

For a cored galaxy, L0 is a stable Lagrange point located at the galactic centre, i.e., $r_\rmG=0$. The nearest saddle point to L0 is L1, located along the x-axis ($\varphi=\varphi_{\rm sep}=0$) at a distance $r_\rmG=r_{\rm sep}=r_1$ from the galactic center.  In the region centred on L0 and bounded by the L1-sparatrix, $\Phi_1$ is bounded by
\begin{align}
q\left[-\frac{1}{\left|r_1 - R\right|}+\frac{r_1}{R^2}\right] < \Phi_1 < q\left[-\frac{1}{R}+\frac{r_1}{2 R^2}\right].
\end{align}
Hence, in the vicinity of L1, the roots of $J_\varphi$ are real for restricted values of $\varphi$ when $E_{\rm Jc}^{'(0)} > h_0^{'(0)} + \Phi_1(r_1,0)$, i.e., when
\begin{align}
E_{\rm Jc}^{'(0)} > E^{'(1)}_\rmJ = h^{'(0)}_0 + q\left[-\frac{1}{\left|r_1 - R\right|} + \frac{r_1}{R^2}\right].
\end{align}
Here $E_{\rm Jc}^{'(0)}$ and $h_0^{'(0)}$ are, respectively, the circular part of the Jacobi energy given by equation~(\ref{Ejcirc_app}) and the unperturbed Jacobi energy, both evaluated at the COP, L0. These trapped orbits that lie inside the L1 separatrix are the \horseshoe orbits. These orbits are therefore in a librating state around near-co-rotation resonances (despite being untrapped wrt L4/L5). The untrapped orbits that lie outside the L1 separatrix, and that satisfy the condition,
\begin{align}
E_{\rm Jc}^{'(0)} < E^{'(1)}_\rmJ,
\end{align}
are the \pacman and the center-phylic orbits. The \pacmans have higher angular momentum ($L^{(1)}<L<L^{(2)}$) than the center-phylic orbits ($L<L^{(1)}$), where $L^{(k)}$, with $k=1,2$, denotes the value of $L$ at the $k^{\rm th}$ Lagrange point. Although the \pacmans are beyond the L1 separatrix, they are still trapped inside the L2 separatrix (as we shall see shortly) and therefore librating about near-co-rotation resonances. The center-phylic orbits on the other hand rotate about the galactic center. Since $J_r$ can vary along an orbit, certain orbits with high $J_r$ can cross the L1 separatrix, resulting in Chimera-like metamorphosis between the trapped \horseshoe and untrapped \pacman and center-phylic orbital families (see Appendix~\ref{App:Chimera}).

\subsection{Perturbation about the perturber}

In the vicinity of the perturber (i.e., the region centred on the perturber and bounded by the L2-separatrix), for a given $r_\rmG$, $\Phi_1$ varies in the range, 
\begin{align}
q\left[-\frac{1}{\left|r_\rmG - R\right|}+\frac{r_\rmG}{R^2}\right]<\Phi_1<q\left[-\frac{1}{R}+\frac{r_\rmG}{2 R^2}\right].
\end{align}
The nearest saddle point to the perturber is L2, located along the x-axis ($\varphi=\varphi_{\rm sep}=0$) at a distance $r_\rmG=r_{\rm sep}=r_2$ from the galactic center. Hence, in the neighborhood of L2, $J_\varphi$ has real roots for restricted values of $\varphi$ when $E_{\rm Jc}^{'\rmP} > h_0^{'\rmP} + \Phi_1(r_2,0)$, i.e., when
\begin{align}
E_{\rm Jc}^{'\rmP} > E^{'(2)}_\rmJ = h^{'\rmP}_0 + q\left[-\frac{1}{\left|r_2 - R\right|}+\frac{r_2}{R^2}\right].
\end{align}
Here $E_{\rm Jc}^{'\rmP}$ and $h_0^{'\rmP}$ are, respectively, the circular part of the Jacobi energy given by equation~(\ref{Ejcirc_app} and the unperturbed Jacobi energy, both evaluated at the COP, which in this case is the perturber. These trapped orbits that lie inside the L2 separatrix are the \pacmans when $E^{'(1)}_\rmJ>E^{'(2)}_\rmJ$ and \horseshoes otherwise. The \pacmans are therefore in a librating state about near-co-rotation resonances, even if they are untrapped wrt L0. The untrapped orbits that lie outside the L2 separatrix, and that satisfy the condition,
\begin{align}
E_{\rm Jc}^{'\rmP} < E^{'(2)}_\rmJ,
\end{align}
are the COM-phylic, perturber-phylic and center-phylic (for $\gamma<0$ profiles) orbits. The COM-phylic orbits have higher angular momentum ($L>L^{(2)}$) than the perturber-phylic orbits ($L^{(1)}<L<L^{(2)}$). While the COM-phylic orbits rotate about the COM of the galaxy-perturber system, the perturber-phylic orbits rotate about the perturber. Due to variation of $J_r$ along an orbit, Chimera-like transitions can occur between the trapped \pacman and untrapped COM-phylic and perturber-phylic orbital families (see Appendix~\ref{App:Chimera}).

\section{Chimera Orbits}
\label{App:Chimera}

\begin{figure*}[t!]
  \centering
  \begin{subfigure}[t]{0.85\textwidth}
    \centering
    \includegraphics[width=1\textwidth]{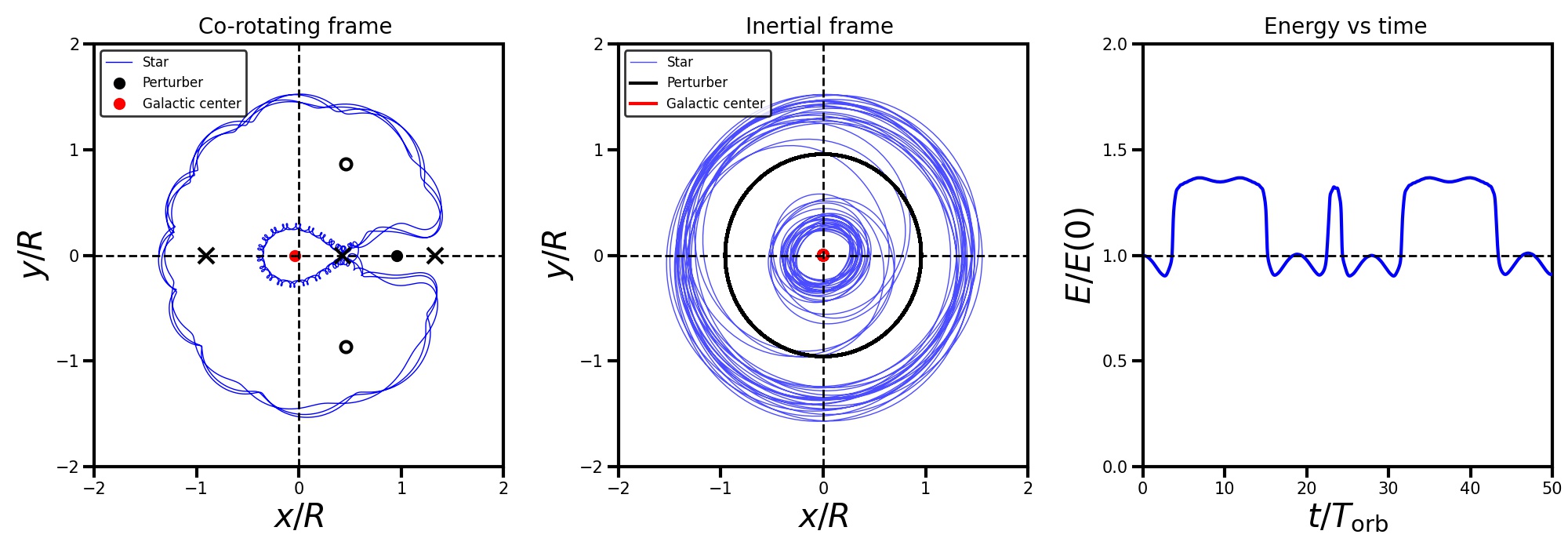}
  \end{subfigure}
  \\
  \begin{subfigure}[t]{0.85\textwidth}
    \centering
    \includegraphics[width=1\textwidth]{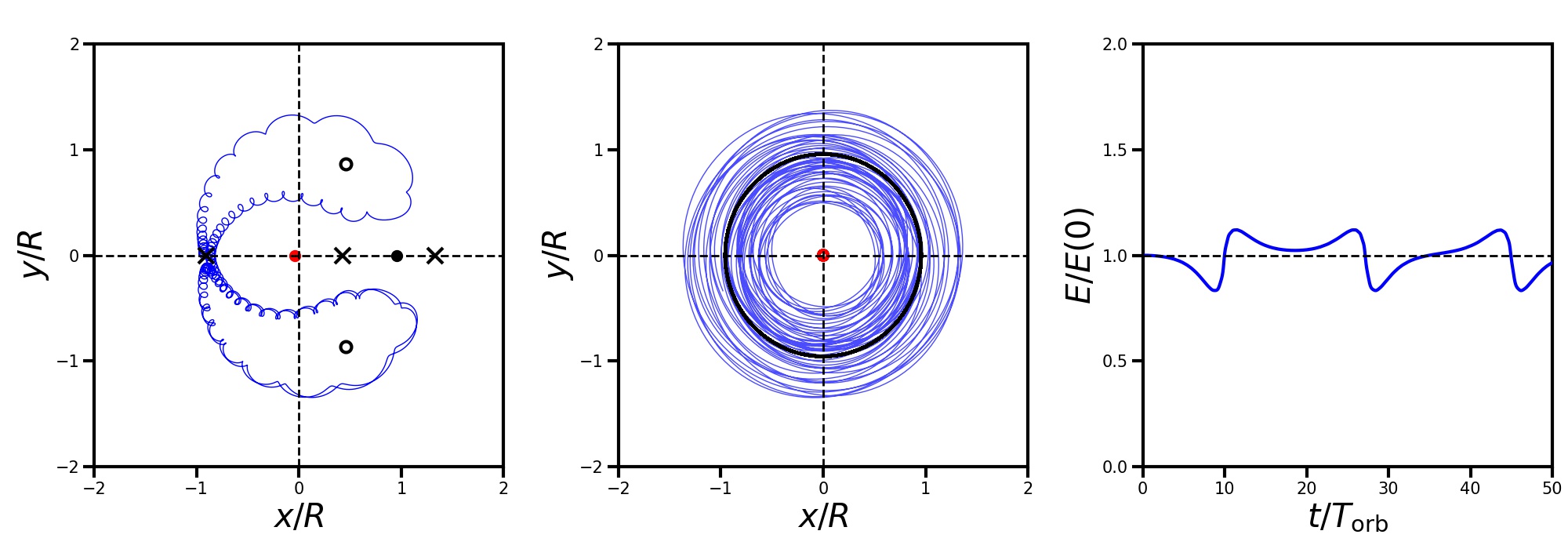}
  \end{subfigure}
  \\
  \begin{subfigure}[t]{0.85\textwidth}
    \centering
    \includegraphics[width=1\textwidth]{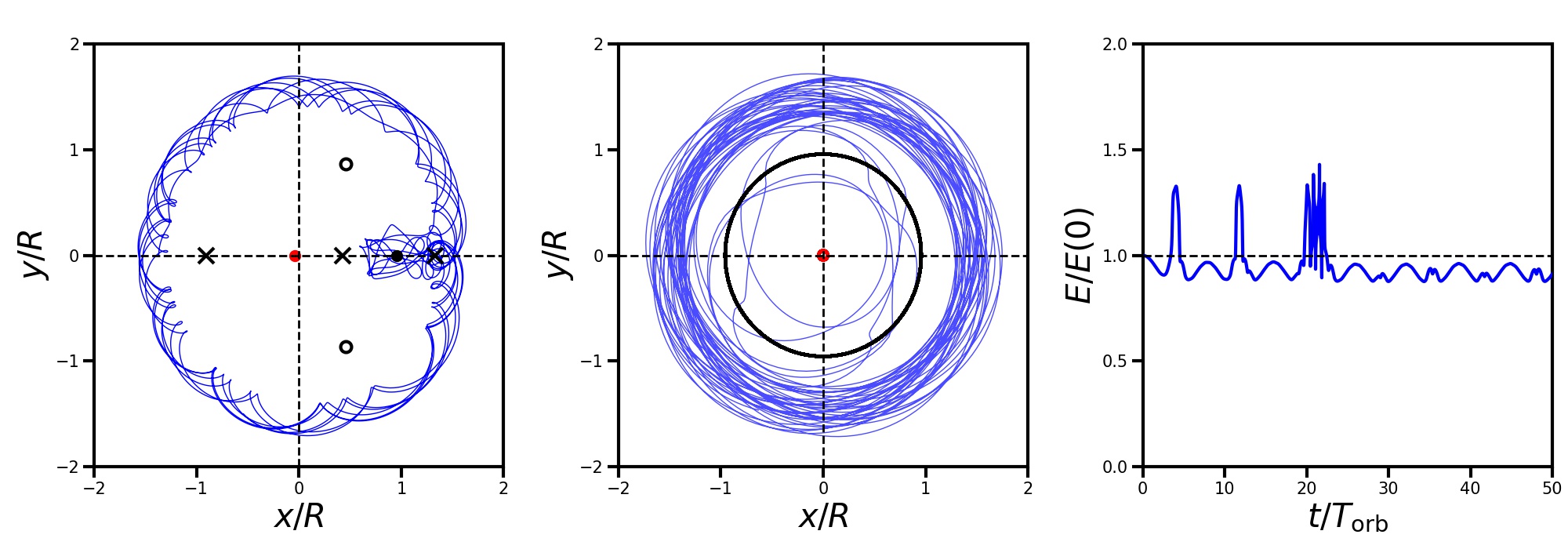}
  \end{subfigure}
  \\
  \begin{subfigure}[t]{0.85\textwidth}
    \centering
    \includegraphics[width=1\textwidth]{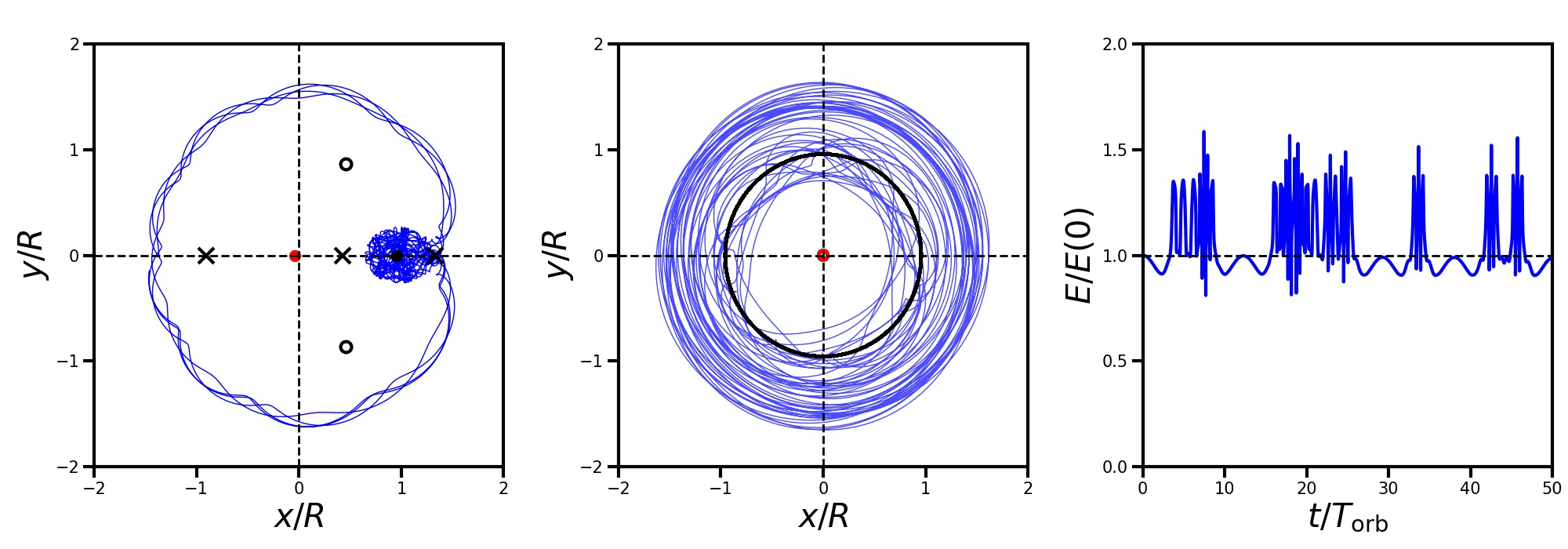}
  \end{subfigure}
  \caption{Examples of Chimera orbits. From top to bottom the panels depict (i) a Chimera orbit initially classified as a \horseshoen, which occasionally undergoes separatrix crossing to transform into a \pacmann, (ii) an initial \horseshoe that transforms into a tadpole, (iii) an initial \pacman that transforms into perturber-phylic and COM-phylic orbits, and (iv) an initial \pacman that occasionally transforms into a perturber-phylic orbit. See text for details.}
  \label{fig:orbit_chimera}
\end{figure*}

A significant subset of the orbits that we classify, based on the circular part of their Jacobi energy (see Appendix~\ref{App:orb_class}), in different orbital families (see Table.~\ref{tab:Ej}), occasionally undergo an inter-family metamorphosis triggered by a separatrix-crossing due to a change in the radial action enabled by the perturber. Fig.~\ref{fig:orbit_chimera} shows a few examples of such Chimera orbits. 

The first row of Fig.~\ref{fig:orbit_chimera} shows a Chimera orbit that is initially classified as a \horseshoe based on the criteria given in Table~\ref{tab:Ej}. However, after taking a detour around L0 like a typical \horseshoen, trapped between the L3 and L1 separatrices, during its first passage along the inner section (part of the orbit inside the perturber's radius), the field particle comes arbitrarily close to L1 during its second passage. Since its ZVC lies very close to the equipotential contour passing through L1, the particle undergoes a separatrix crossing (L1 separatrix) after which it takes a shortcut in between L0 and the perturber and becomes a \pacman orbit (based on the criteria given in Table~\ref{tab:Ej}), trapped between the L1 and L2 separatrices. After behaving like a \pacman during its second passage, the particle crosses the L1 separatrix again during its third passage to re-enter the \horseshoe phase. These \horseshoe $\rightarrow$ \pacman $\rightarrow$ \horseshoe transformations of the Chimera orbit are evident from the energy curve (right-hand panel), where a short-period oscillation corresponding to the \pacman phase is sandwiched between two long-period oscillations corresponding to the \horseshoe phases (cf. top and middle rows of Fig.~\ref{fig:orbit2}).

The second row depicts a Chimera orbit that is initially classified as a \horseshoe (trapped between the L3 and L1 separatrices), but which transforms into a tadpole (trapped within the L3 separatrix). In its \horseshoe phase, the particle makes a full circulation around L4 and L5 and its energy undergoes a long period oscillation (see right-hand panel). Then it enters its tadpole phase where it circulates only L5 and its energy undergoes a short period oscillation with a period exactly half of that of its \horseshoe phase. The separatrix-crossing in this case is triggered when the particle comes arbitrarily close to L3. This particular orbit has a ZVC that lies close to the L3 separatrix, which is why both the \horseshoe and tadpole phases have very long libration periods ($T_{\rm lib}$ asymptotes to infinity as the particle approaches the separatrix).

The third row shows a Chimera orbit that is initially classified as a \pacman orbit (trapped between the L1 and L2 separatrices), but which transforms into a perturber-phylic and a COM-phylic orbit, both of which lie beyond the L2 separatrix. In its initial \pacman phase, the particle undergoes regular, long period oscillations in energy (see right-hand panel). Then it comes arbitrarily close to L2 and undergoes a separatrix-crossing (L2 separatrix) to enter the perturber-phylic phase, which is reminiscent of resonant capture. In this phase the particle rotates around the perturber, associated with rapid oscillations in energy. At some point the particle approaches L2 again and enters a COM-phylic phase associated with energy oscillations that have much smaller amplitude than those during the \pacman and perturber-phylic phases.

Finally, the fourth row depicts a Chimera orbit that is initially classified as a \pacman orbit, trapped between the L1 and L2 separatrices, but which undergoes frequent L2 separatrix-crossings to become perturber-phylic. Note how the regular, long-period oscillations in energy corresponding to its \pacman phase are interspersed with rapid oscillations corresponding to its perturber-phylic phase. 

\label{lastpage}

\end{document}